\documentclass[
  reprint,
  amsmath,amssymb,
  aps,
  prx,
  superscriptaddress,
  floatfix
]{revtex4-2}

\usepackage{graphicx}      % 插图支持
\usepackage{dcolumn}       % 小数点对齐表格列
\usepackage{bm}            % 数学粗体符号，例如 \bm{k}
\usepackage{mathtools}     % 补充数学功能
\usepackage{braket}        % Dirac 表示
\usepackage{subfigure}     % 并列图
\usepackage{booktabs}      % 高质量表格线
\usepackage{multirow}      % 表格多行合并
\usepackage{xcolor}        % 文字颜色
\usepackage{hyperref}      % 超链接支持
\usepackage{placeins}
\usepackage{array}      % 增强列格式
\usepackage{makecell}

\hypersetup{
    colorlinks=true,
    linkcolor=blue,
    citecolor=blue,
    urlcolor=blue
}
\begin{document}

\title{Electronic Toroidal Metals: Landau Theory and Magnetoelectric Fingerprints}

\author{Yuhang Xiao}
\affiliation{Anhui Province Key Laboratory of Low-Energy Quantum Materials and
Devices, High Magnetic Field Laboratory, HFIPS, Chinese Academy
of Sciences, Hefei, Anhui 230031, China}
\affiliation{Science Island Branch of Graduate School, University of Science and
	Technology of China, Hefei, Anhui 230026, China}
\author{Ning Hao}
\email{haon@hmfl.ac.cn}
\affiliation{Anhui Province Key Laboratory of Low-Energy Quantum Materials and
Devices, High Magnetic Field Laboratory, HFIPS, Chinese Academy
of Sciences, Hefei, Anhui 230031, China}

\begin{abstract}

Toroidal order is a higher-rank multipolar order whose intrinsic realization in itinerant-electron systems remains unexplored. Here, we develop a generic theory of the electronic toroidal metal (ETM), in which toroidal order emerges spontaneously from electronic degrees of freedom near the Fermi surfaces. Because candidate toroidal bilinears can overlap by symmetry with the uniform charge current, we formulate a projected instability criterion that removes the noncondensable current component. Within this current-orthogonal sector, we identify a soft collective mode in the $\mathcal{P}$-odd and $\mathcal{T}$-odd particle-hole channel and demonstrate that ETM arises as a Fermi-liquid instability. We then define the toroidal moment in an itinerant system through the antisymmetric magnetoelectric response tensor. We further establish an intimate connection between this response and the topology of pseudospin texture: the rearrangement of pseudospin vortices changes the winding structure of the Fermi surfaces and markedly enhances the magnetoelectric response. ETM also exhibits characteristic nonlinear charge transport, including a pronounced enhancement of its interband quantum-geometric contribution. Our results establish ETM as a distinct metallic Landau phase and provide a general framework for understanding toroidal order generated by itinerant electrons.

\end{abstract}

\maketitle

% === main text begins ===
\section{introduction}

Beyond conventional Landau orders characterized by charge, polarization, and magnetization, higher-rank multipolar orders provide a broader framework for describing symmetry breaking in quantum materials\cite{RevModPhys.81.807,Pourovskii2025,PhysRevB.64.195109,Chandra_2002,PhysRevLett.105.157003,PhysRevLett.115.026401,Chu_2010,Kuo_2016,Ronning_2017}. Among them, toroidal order is particularly distinctive. The toroidal moment arises in the multipole expansion of a current distribution at the same order as the magnetic quadrupole moment, but, unlike an ordinary magnetic multipole, it doesn't produce magnetostatic field outside the source region and instead couples to the curl of the magnetic field\cite{Landau1980,Dubovik_1990}. It is odd under both spatial inversion $\mathcal P$ and time reversal $\mathcal T$, while remaining even under their product $\mathcal{PT}$, and therefore transforms in the same symmetry channel as the antisymmetric part of the linear magnetoelectric tensor\cite{Spaldin_2008,PhysRevLett.102.157203,Van_Aken_2007,PhysRevB.82.100408,Zimmermann_2014,Ding_2021}.

The microscopic description of toroidal order depends crucially on the nature of the underlying electronic degrees of freedom. For systems composed of localized magnetic moments, the natural real-space expression $\bm{T}\sim\sum_i\bm{r}_i\times\bm{m}_i$ is sensitive to the choice of origin and crystallographic unit cell. In insulating magnets, this ambiguity can be treated in close analogy with the modern theory of polarization: toroidization is a multivalued bulk quantity, and only its changes under adiabatic deformations are uniquely defined\cite{PhysRevB.76.214404}. Further, a semiclassical formulation introduces a thermodynamic response tensor $\mathcal{Q}_{ij}$ through $\delta F\sim \mathcal{Q}_{ij}\partial_iB_j$ and identifies its antisymmetric component with the toroidal moment\cite{PhysRevB.97.134423,PhysRevB.98.060402}. These approaches establish how toroidal order can be defined and characterized when it originates from localized moments, or when itinerant carriers experience an essentially static toroidal background. The problem addressed here is complementary: we ask whether the toroidal order parameter itself can arise dynamically from itinerant electronic degrees of freedom near the Fermi surfaces (FS).

This itinerant problem is qualitatively different from its localized-moment counterpart. In a metal, the occupied states do not form an isolated manifold separated by an excitation gap, so the localized-Wannier and bulk-polarization constructions appropriate to insulators cannot be transferred directly. Moreover, static electric fields are efficiently screened in metals, and the low-energy electronic response is governed by particle–hole excitations in the vicinity of the FS. The central question is therefore not whether metallic carriers can coexist with a preexisting toroidal texture, but whether a particle-hole mode that is odd under both $\mathcal P$ and $\mathcal T$, yet even under $\mathcal{PT}$, can soften and condense within an interacting Fermi liquid. We refer to the resulting phase as an electronic toroidal metal (ETM), and denote its Landau order parameter by the toroidal vector $\bm{\eta}$.

A fundamental obstruction arises in formulating this instability. Electronic bilinears that transform as a toroidal vector have exactly the same spacetime symmetry as a uniform electric current. Symmetry alone therefore cannot distinguish a genuine toroidal order parameter from a current-carrying deformation of the Fermi sea. The latter cannot condense in thermodynamic equilibrium, because a spatially homogeneous equilibrium state must carry no net current\cite{Bachmann_2020,PhysRevLett.123.060601}, a constraint ultimately enforced by electromagnetic gauge invariance. The uniform-current component must consequently be removed before the physical Landau instability is identified. We implement this removal through a Schur-complement projection defined by the susceptibility matrix. The resulting Schur kernel selects the genuinely condensable collective mode in the subspace orthogonal to the current sector and simultaneously generates a backflow correction to its coupling with the electrons. This projection determines not only the instability criterion, but also the effective electron-boson vertex and the ordered-state mean-field Hamiltonian.

The itinerant origin of the order also calls for an experimentally meaningful definition of the toroidal moment. We define the metallic toroidal moment $\bm{T}$ through the antisymmetric component of the magnetoelectric response tensor, while reserving $\bm{\eta}$ for the spontaneously condensed toroidal vector. In localized-moment theories, toroidal order is commonly treated as a static microscopic arrangement. Although such an arrangement naturally permits an antisymmetric magnetoelectric response, a measurement at a fixed point in the ordered phase does not by itself establish that the observed response originates uniquely from toroidal order, because other symmetry-equivalent order parameters such as $\bm{P}\times\bm{M}$ can produce the same tensor component\cite{PhysRevLett.95.237402,PhysRevB.87.014421,PhysRevLett.106.057403,Park_2022}. The ETM framework provides additional information by describing the complete phase transition: the softening of the collective mode, the onset of $\bm{\eta}$, and the accompanying evolution of $\bm{T}$ can all be followed continuously across the transition. The temperature dependence therefore provides a substantially sharper experimental fingerprint than the mere presence of an antisymmetric magnetoelectric coefficient.

We demonstrate this framework using a minimal spin-orbit-coupled lattice model that preserves both $\mathcal P$ and $\mathcal T$ in the normal state. The leading eigenmode of the Schur kernel condenses into a $\mathcal{PT}$-symmetric ETM. The twofold degeneracy of each band remains protected, while the two FS are displaced relative to one another in momentum space. The toroidal vector $\bm{\eta}$ also continuously reconstructs the pseudospin texture: as $|\bm{\eta}|$ develops, pseudospin vortices move through the Brillouin zone (BZ), and when they cross the FS, the winding numbers on the inner and outer FS change accordingly. This topological rearrangement produces a pronounced enhancement of the intrinsic interband magnetoelectric response and hence of the response-defined toroidal moment $\bm{T}$. 

Beyond linear magnetoelectricity, the ETM exhibits distinctive nonlinear charge transport. Among the four representative spacetime-symmetry classes of metallic Landau orders, the ETM is distinguished by the presence of an interband quantum-metric-dipole contribution\cite{PhysRevB.108.L201405,4z8z-4kch,shibata2026}. Moreover, the same pseudospin-vortex evolution that enhances $\bm{T}$ produces a pronounced enhancement of this interband quantum-geometric response. Taken together, the Schur-projected Fermi-liquid instability, the transition-resolved toroidal moment, and the intrinsic magnetoelectric and transport responses establish the ETM as a distinct metallic Landau phase and provide a general framework for toroidal order generated by itinerant electrons.

\section{Theoretical framework of electronic toroidal metal}

A general theory of ETM begins by identifying electronic bilinears that transform as toroidal order parameters and determining which of these channels is susceptible to a low-energy instability. Because the ETM phase arises from a $\mathcal{P}$-odd and $\mathcal{T}$-odd instability of an interacting Fermi liquid, the relevant particle–hole bilinears must be odd under both spatial inversion and time reversal. These symmetry-allowed bilinears span the candidate order-parameter space on which the subsequent Landau instability analysis is built.

A finite set of such bilinears is constructed by expanding the electronic bilinears in lattice harmonics, or equivalently in powers of momentum in the long-wavelength limit. The resulting operators, which we refer to as toroidal candidates, are classified according to their crystalline symmetry and serve as the basis for the instability analysis:
\begin{equation}
    \hat{\Gamma}_i^\mu(\bm{q}) = \sum_{\bm{k}} c_{\bm{k}+\bm{q}/2}^\dagger \Gamma_i^\mu(\bm{k}) c_{\bm{k}-\bm{q}/2}.\label{formal_toroidal_candidates}
\end{equation}
Here $i$ denotes the spatial component, while $\mu$ labels distinct symmetry channels. Electromagnetic gauge invariance precludes a uniform electric current from condensing as an equilibrium order parameter. This constraint is particularly consequential for ETM because the toroidal candidates and the electric current transform identically under space-time symmetries. Consequently, the toroidal candidates may include operators that either coincide with the uniform current or dynamically mix with it. This issue does not arise for more conventional Landau orders, such as spin or charge-density order, whose space-time transformation properties differ from those of the current and therefore forbid such mixing.

We must therefore remove the uniform-current component from the toroidal candidates in an effective manner, and determine the softening instability only within the subspace that is orthogonal to the uniform-current channel in the sense of the static susceptibility. Technically, this projection may be formulated by using the correlation function $\Pi_{AB}$ between two electronic bilinears to define an inner product, $\langle A,B\rangle=\Pi_{AB}(i\Omega\to 0,\bm q\to 0)$, and then orthogonalizing the space spanned by all toroidal candidates together with the current operators. In what follows, however, we shall implement the same Schur projection equivalently within the path-integral formulation, where the underlying physical content of the procedure is more transparent.

\subsection{Schur-projection procedure}

We consider the action $S[\bar{c}, c, \phi, a] = S_0[\bar{c}, c] + S_{\text{aux}}[\phi, a] + S_{\text{int}}[\bar{c}, c, \phi, a]$, where $S_0$ describes Bloch electrons, while $S_{\text{aux}}$ and $S_{\text{int}}$ denote, respectively, the bare action of the auxiliary fields and their coupling to electrons. Their explicit forms are given by:
\begin{align}
&S_{\text{aux}} = \frac{1}{2} \sum_q \phi_i^\mu(-q) [g_{\rm T}^{-1}]_{ij}^{\mu\nu} \phi_j^\nu(q),\label{S_aux}\\
&S_{\text{int}} = \sum_{k,q} \bar{c}_{k+q/2} [\phi_i^\mu(q) \Gamma_i^\mu(\bm{k}) + a_i(q) v_i(\bm{k})] c_{k-q/2},\label{S_int}
\end{align}
where $q=(i\Omega_n,\bm{q})$, $v_i$ is the velocity of Bloch electrons and $g_{\rm T}$ is the coupling tensor in toroidal channel. For simplicity, we take an isotropic coupling, $[g_{\rm T}^{-1}]^{\mu\nu}_{ij} =g_{\rm T}^{-1}\mathbf{1} $. In addition to the auxiliary fields $\phi^{\mu}_i$ associated with the toroidal candidates, we introduce an auxiliary field $\bm{a}$ coupled to the electric current, which allows us to implement the following Schur projection. It is important to emphasize, however, that $\bm{a}$ is only introduced as a mathematical device: it carries no bare coupling and dynamics and should not be identified with the physical electromagnetic gauge field.

After integrating out the electrons, we obtain the effective quadratic action as follows:
\begin{align}
&S_{\text{eff}} = \frac{1}{2} \sum_q \left( \phi_i^{\mu}(-q) \,\,\,\, a_i(-q) \right) 
K(q)
\begin{pmatrix}
\phi_j^{\nu}(q) \\
a_j(q)
\end{pmatrix},\label{S_eff}\\
&K(q)=\begin{pmatrix}
 g_{\rm T}^{-1} + \Pi_{i\mu,j\nu}^{\Gamma\Gamma}(q) & \Pi_{i\mu,j}^{\Gamma J}(q) \\[5pt]
\Pi_{i,j\nu}^{J\Gamma}(q) &  \Pi_{ij}^{JJ}(q)
\end{pmatrix},
\end{align}
where we have defined the following correlation functions:
\begin{align}
&\Pi_{i\mu,j\nu}^{\Gamma\Gamma}(q) = \sum_k\operatorname{Tr} \left[ G_0(k_+)\Gamma_i^{\mu}(\bm{k})G_0(k_-)\Gamma_j^{\nu}(\bm{k}) \right],\label{Pi_GammaGamma}\\
&\Pi_{i\mu,j}^{\Gamma J}(q) = \sum_k \operatorname{Tr} \left[ G_0(k_+)\Gamma_i^{\mu}(\bm{k})G_0(k_-)v_j(\bm{k}) \right],\label{Pi_GammaJ}\\
&\Pi_{ij}^{JJ}(q) = \sum_k \operatorname{Tr} \left[ G_0(k_+)v_i(\bm{k})G_0(k_-)v_j(\bm{k}) \right],\label{Pi_JJ}
\end{align}
with $\sum_k= T \sum_{\omega_n} \int_{\rm BZ}\mathrm{d}\bm{k}$ and $k_{\pm}=k\pm q/2$. Here $G_0$ is the Bloch-electron propagator. We then integrate out the auxiliary field $\bm{a}$. This step is essential: mathematically, it is equivalent to performing the Schur projection in the inner-product space introduced above. Physically, because the current and the toroidal candidates share the same symmetry, their dynamical mixing is unavoidable, as reflected in the off-diagonal components of the $K(q)$ matrix. Fluctuations of the toroidal candidates therefore induce current fluctuations, while the latter in turn generate a backflow correction that renormalizes the former. One must decouple these two sectors dynamically, so that the projected toroidal candidates no longer contain the current component. As we shall see, this procedure also naturally modifies the way in which the toroidal candidates couple to the electrons. This provides an interesting example in which dynamically projecting out part of the bosonic degrees of freedom profoundly reshapes the coupling between the remaining bosonic modes and the underlying electrons. 

After carrying out the Gaussian integration over the auxiliary field $\bm{a}$ , we obtain an effective action involving only the toroidal candidates as $S_{\text{eff}}[\phi] \sim\sum_q \phi(-q) \left(g_{\rm T}^{-1}-\Pi^{\text{Schur}}(q)\right) \phi(q)$, where the Schur-kernel matrix reads:
\begin{equation}
\Pi^{\text{Schur}}(q)= -\Pi^{\Gamma\Gamma}(q) + \Pi^{\Gamma J}(q) \left[ \Pi^{JJ}(q) \right]^{-1} \Pi^{J\Gamma}(q).\label{Schur_kernel}
\end{equation}
Here the indices have been suppressed. The genuinely condensable bosonic mode, is therefore selected by the eigenvector associated with the largest positive eigenvalue of the Schur kernel in the $q\to 0$ limit. Correspondingly, the transition temperature $T_{\rm c}$ is determined by $1-g_{\rm T}\lambda_{\max}(T_{\rm c})=0$, where $\lambda_{\max}(T)$ denotes the largest eigenvalue of $\Pi^{\rm Schur}(T)\equiv\Pi^{\rm Schur}(q\to 0,T)$.

The Schur kernel Eq.~\ref{Schur_kernel} is the central result of the Schur projection. It consists of two contributions: the correlation function in the bare toroidal-candidate channel and the backflow correction $\Pi^{\Gamma J}  \left(\Pi^{JJ}\right)^{-1} \Pi^{J\Gamma}$ generated by the current channel. Its mathematical structure shows that, if a given toroidal candidate has a strong dynamical overlap with the current, its effective kernel is strongly renormalized. By contrast, if a candidate is completely orthogonal to the current, namely $ \Pi_{i\mu,j}^{\Gamma J}(q)=0 $, then the current channel produces no backflow correction to it. And it should be emphasized that the current-current correlation function $\Pi^{JJ}$ contains only the paramagnetic contribution, rather than the full electromagnetic response kernel. Consequently, taking its inverse in the $q\to 0$ limit is nonsingular.

\subsection{Effective electron-boson coupling}

In the course of carrying out the Gaussian integration over the auxiliary field $\bm a$, one can extract its saddle-point configuration $\bm{a}^{\rm sp}$. It is determined by the saddle-point equation $\delta S_{\rm eff}[\phi,\bm{a}]/\delta a_i=0$, and may be interpreted as the current fluctuation induced by a given toroidal-candidate fluctuation. Explicitly, it takes the form:
\begin{equation}
a_i^{\text{sp}}(q) = -[\Pi^{JJ}(q)^{-1}]_{ij} \Pi_{j,m\mu}^{J\Gamma}(q) \phi_m^\mu(q).
\end{equation}
Substituting it back into the bare electron-boson coupling vertex yields the effective interaction as follows:
\begin{align}
&S_{\text{int}}^{\text{eff}} = \sum_{k,q}\tilde\Gamma_i^{\mu}(\bm{k},q) \phi_i^\mu(q)\bar{c}_{k+q/2}     c_{k-q/2},\label{effective_interaction}\\
&\tilde{\Gamma}_i^{\mu}(\bm{k},q) = \Gamma_i^{\mu}(\bm{k}) - v_m(\bm{k}) \left[  \Pi^{JJ}(q) \right]_{mj}^{-1} \Pi_{j,i\mu}^{J\Gamma}(q).\label{effective_vertex}
\end{align}
Starting from this effective interaction and integrating out the electrons, one can obtain the Schur kernel Eq.~(\ref{Schur_kernel}) again. The effective interaction is therefore consistent with the Schur-projection procedure implemented above.
We note that the bare vertex depends only on the electronic momentum $\bm k$, and therefore represents a local and instantaneous coupling. After the backflow correction is incorporated, however, the effective vertex acquires an explicit dependence on the bosonic momentum and frequency. It thus describes a nonlocal and retarded coupling, endowing the electron-boson problem with richer dynamical structure. Consequently, the resulting electronic self-energy cannot, in general, be reduced to a simple renormalization of the band dispersion. 

The mean-field Hamiltonian describing the ETM is obtained by taking the $q\to 0$ limit of Eq.~(\ref{effective_interaction}) and substituting the softened eigenvector of the Schur matrix. Owing to the backflow correction, the resulting mean-field Hamiltonian takes a form fundamentally different from that of other Landau orders in metals. As we shall see, its distinctive spectroscopic, electromagnetic-response, and transport signatures provide diagnostic features of the ETM.

\section{Observable Fingerprints}

To place the formal construction above on concrete footing, we consider a minimal two-band Hamiltonian on a square lattice as a prototype for spin-orbit-coupled systems preserving both $\mathcal P$ and $\mathcal T$ symmetries\cite{Samokhin_2022,PhysRevLett.108.147003,PhysRevResearch.3.023204}. The Bloch Hamiltonian reads:
\begin{equation}
H_0(\bm{k}) = \xi(\bm{k}) \tau_0 \sigma_0 + m(\bm{k}) \tau_x \sigma_0 + g_x(\bm{k}) \tau_z \sigma_x + g_y(\bm{k}) \tau_z \sigma_y.\label{H0}
\end{equation}
The momentum functions are $\xi(\bm{k}) = -2t(\cos k_x + \cos k_y) - \mu, \quad m(\bm{k})=m_0$, $g_x(\bm{k}) = -\lambda_{\rm SOC} \sin k_y$ and $g_y(\bm{k}) = \lambda_{\rm SOC} \sin k_x$, where $t$, $\mu$ and $\lambda_{\rm SOC}$ are hopping parameter, chemical potential and SOC strength, respectively. The $\sigma$ and $\tau$ Pauli matrices act on the spin and orbital degrees of freedom, respectively, and the time-reversal and inversion operators are given by $\mathcal T=i\sigma_y K$ and $\mathcal P=\tau_x$. Up to nearest-neighbor coupling, the kernels of the toroidal-candidate operators selected by symmetry are listed in Table \ref{tab:toroidal_candidates}. In the present minimal model, we focus on the two-dimensional representation of the $C_{4h}$ group, corresponding to in-plane toroidal fluctuations. This restricted setting is already sufficient to illustrate the formal construction developed above, to predict a variety of observable signatures, and, in particular, to reveal the distinctive effects arising from the topological pseudospin structures in momentum space.

\begin{table}
\caption{Five sets of toroidal candidates belonging to the $E_u$ representation of the $C_{4h}$ group.}
\label{tab:toroidal_candidates}
\centering
% --- column spacing ---
\setlength{\tabcolsep}{8pt}
% --- row spacing ---
\renewcommand{\arraystretch}{1.3}
\begin{tabular}{c c c}
\hline
$\mu$ & $\Gamma_x^\mu(\mathbf{k})$ & $\Gamma_y^\mu(\mathbf{k})$ \\ \hline
1    & $\tau_z \sigma_y$          & $-\tau_z \sigma_x$          \\
2    & $(\cos k_x + \cos k_y)\tau_z \sigma_y$ & $-(\cos k_x + \cos k_y)\tau_z \sigma_x$ \\
3    & $(\cos k_x - \cos k_y)\tau_z \sigma_y$ & $(\cos k_x - \cos k_y)\tau_z \sigma_x$ \\
4    & $\sin k_x \, \tau_0 \sigma_0$ & $\sin k_y \, \tau_0 \sigma_0$ \\
5    & $\sin k_x \, \tau_x \sigma_0$ & $\sin k_y \, \tau_x \sigma_0$ \\ \hline
\end{tabular}
\end{table}

\begin{figure}
\includegraphics[width=0.96\columnwidth]{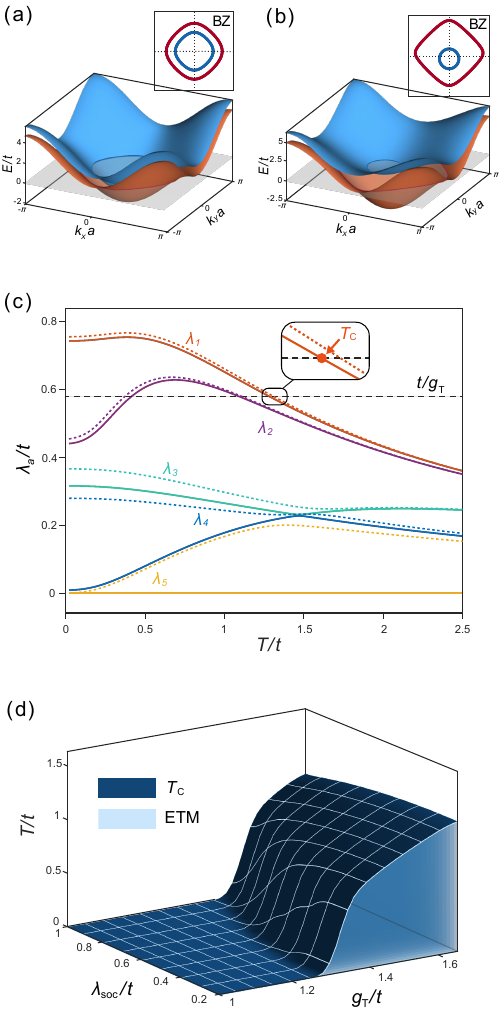}
\caption{(a),(b) Band structures and FS before and after condensation ($\theta=\pi/2$), respectively. (c) Temperature dependence of the Schur-kernel eigenvalues, shown as solid curves. The corresponding eigenvalues obtained without the Schur projection are shown as dashed curves for comparison. The transition temperature is determined by the intersection between $t/g_{\rm T}$ and the largest eigenvalue branch of the Schur kernel. (d) Phase diagram of the ETM. The parameters used in panels (a)--(c) are $\mu=-1.4t$, $m_0=0.5t$, and $\lambda_{\rm SOC}=0.3t$, which are also adopted in all relevant numerical calculations of this model presented thereafter.
 }
\label{fig1}
\end{figure}

\subsection{Spectroscopic features}
\subsubsection{Fermi-surface distortion}
We calculate all correlation functions, whose analytical expressions are provided in the Supplemental Material, and assemble them into the Schur kernel using Eq.~(\ref{Schur_kernel}). Owing to the $C_{4h}$ symmetry, each eigenvalue in the toroidal $E_u$ representation is twofold degenerate. In the continuum limit, this degenerate manifold corresponds to rotations of the in-plane Goldstone mode. In Fig.~\ref{fig1}(c), we show the temperature dependence of the five eigenvalue branches of the Schur kernel and compare them with the corresponding eigenvalues obtained without the Schur projection. As the temperature is lowered, the correction induced by the Schur projection becomes increasingly pronounced. For the present model, omitting the Schur projection only slightly overestimates the transition temperature $T_{\rm c}$. However, the two results are qualitatively different: after the Schur projection, the lowest eigenvalue branch is pinned to zero, and the corresponding subspace is precisely the uniform-current sector. If the toroidal candidate operators have a stronger dynamical overlap with the uniform current, the correction of $T_{\rm c}$ induced by the Schur projection can become substantially more significant.

By numerically calculating the transition temperature over a range of SOC strengths and toroidal coupling, we obtain the phase diagram of the ETM as shown in Fig.~\ref{fig1}(d). In the weak- to intermediate-SOC regime, the emergence of the ETM requires at least a moderate bare interaction in the toroidal channel.

The mean-field Hamiltonian obtained after the softening of the leading eigenmode takes the form:
\begin{align}
    H_{\text{MF}}(\bm{k}) = H_0(\bm{k}) &+ \sum_{a}\eta_a \sum_{A=1}^{10} e^a_A \Big[ \Gamma_A(\bm{k}) \nonumber\\
    &-\sum_{m,n} v_m(\bm{k}) \left( \Pi^{JJ} \right)^{-1}_{mn} \Pi_{nA}^{J\Gamma}\Big],\label{H_MF}
\end{align}
where $A=(i,\mu)$ is a composite index running over the 10 toroidal candidates, and $e_A^a$ are the components of the leading eigenmode. All the correlation functions in Eq.~(\ref{H_MF}) are taken in the limits $q\to 0$ and $T\to 0$. Here $\bm{\eta}=(\eta_x,\eta_y)=\eta(\cos\theta,\sin\theta)$ denotes the in-plane condensate vector. We identify $\bm{\eta}$ as the in-plane toroidal vector, which transforms as $\bm{\eta}\to-\bm{\eta}$ under either $\mathcal{T}$ or $\mathcal{P}$, while remaining invariant under $\mathcal{PT}$.

\begin{table*}[t]
\caption{
Metallic Landau orders of itinerant-electron origin in four distinct spacetime-symmetry sectors. We consider a normal metal with $C_{4h}$ symmetry and choose one representative order parameter in each sector, together with its characteristic electromagnetic responses. Different order-parameter choices within the same sector can give rise to different responses. For instance, in the magnetic-quadrupolar metal considered here, the spin polarization is taken to be out of plane, and the resulting $C_4\mathcal{T}$ symmetry forbids a linear anomalous Hall response.
}
\label{tab:PT_classification}
\centering
\renewcommand{\arraystretch}{1.4}
\setlength{\tabcolsep}{4pt}

\begin{ruledtabular}
\begin{tabular}{ccccc}
$(\mathcal{P},\mathcal{T})$
&
\parbox[c]{0.15\textwidth}{\centering Ordered phase}
&
\parbox[c]{0.20\textwidth}{\centering Order parameter}
&
\parbox[c]{0.10\textwidth}{\centering FS}
&
\parbox[c]{0.31\textwidth}{\centering Electromagnetic response}
\\
\hline\\[-6pt]
$(-,-)$
&
\parbox[c]{0.15\textwidth}{\centering ETM}
&
\parbox[c]{0.20\textwidth}{\centering requires Schur projection}
&
\parbox[c]{0.10\textwidth}{\centering Fig.~\ref{fig2}(a)}
&
\parbox[c]{0.31\textwidth}{\centering magnetoelectric response /\\ nonlinear charge transport}
\\[15pt]
$(+,+)$
&
\parbox[c]{0.15\textwidth}{\centering charge-nematic metal}
&
\parbox[c]{0.20\textwidth}{\centering $(\cos k_x-\cos k_y)\tau_0\sigma_0$}
&
\parbox[c]{0.10\textwidth}{\centering Fig.~\ref{fig2}(b)}
&
\parbox[c]{0.31\textwidth}{\centering anisotropic resistivity /\\elastoresistive effect}
\\[15pt]
$(-,+)$
&
\parbox[c]{0.15\textwidth}{\centering polar metal}
&
\parbox[c]{0.20\textwidth}{\centering $\tau_0(\sin k_y\sigma_x-\sin k_x\sigma_y)$}
&
\parbox[c]{0.10\textwidth}{Fig.~\ref{fig2}(c)}
&
\parbox[c]{0.31\textwidth}{Edelstein effect /\\inverse Edelstein effect /\\nonlinear Hall effect}
\\[15pt]
$(+,-)$
&
\parbox[c]{0.15\textwidth}{\centering magnetic quadrupolar metal}
&
\parbox[c]{0.20\textwidth}{\centering $(\cos k_x-\cos k_y)\tau_0\sigma_z$}
&
\parbox[c]{0.10\textwidth}{\centering Fig.~\ref{fig2}(d)}
&
\parbox[c]{0.31\textwidth}{\centering crystal spin Hall effect /\\magneto-optical Kerr effect}
\\[6pt]
\end{tabular}
\end{ruledtabular}
\end{table*}

\begin{figure*}
\centering
\includegraphics[width=0.88\textwidth]{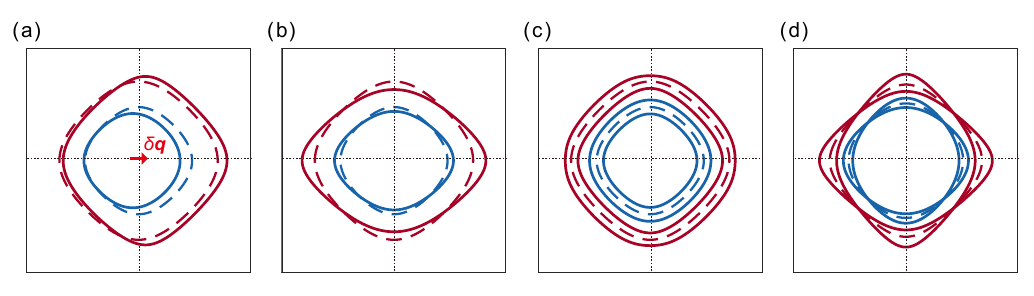}
\caption{Schematic illustration of the characteristic Fermi-surface deformations in metallic Landau phases belonging to four distinct spacetime-symmetry sectors. Dashed curves indicate the FS in the normal metallic state, while solid curves indicate those after order-parameter condensation.
}
\label{fig2}
\end{figure*}

The band structures and FS of the normal metal before condensation and of the resulting ETM phase are shown in Figs.~\ref{fig1}(a) and \ref{fig1}(b), respectively. Due to $\mathcal{PT}$ symmetry, the two-fold degeneracy of each band is preserved. A salient feature of the ETM FS is that, although the geometry of each individual FS remains essentially unchanged, the two FS are shifted relative to each other in momentum space, leaving $\mathcal{M}_\theta$--the mirror symmetry dictated by the orientation angle $\theta$--as the sole remaining point-group symmetry. The corresponding displacement vector $\delta\bm{q}$ is controlled by the toroidal vector $\bm{\eta}$ and can therefore be tuned continuously by temperature in ETM phase.

For comparison, we summarize representative metallic Landau orders in the other three spacetime-symmetry sectors in Table \ref{tab:PT_classification}. Their characteristic Fermi-surface distortions in the corresponding ordered phases are shown in Fig.~\ref{fig2}. In the charge-nematic metal\cite{PhysRevB.64.195109,PhysRevB.70.155110,Borzi_2007,Lawler_2010}, the rotational symmetry of the FS is reduced from $C_4$ to $C_2$. In the polar metal\cite{PhysRevLett.14.217,PhysRevLett.124.237601,PhysRevB.107.165110,Fei_2018}, the $C_4$ symmetry is preserved, but the originally twofold-degenerate bands are split by the polar order. In the magnetic-quadrupolar metal\cite{PhysRevB.102.014422,PhysRevX.12.031042,PhysRevLett.132.176702,Reichlova_2024,Jiang_2025}, the band degeneracy is lifted and the FS acquire a $d$-wave-like distortion. A common feature of these three metallic Landau orders is that the geometric centers of the Fermi pockets are not displaced relative to each other. This sharply contrasts with the ETM phase, where the primary Fermi-surface signature is precisely the relative momentum-space displacement between the two $\mathcal{PT}$-protected FS.

\subsubsection{Topological structure in momentum space}

Beyond Fermi-surface distortion, the conserved quantum number provides another useful diagnostic. In ETM, owing to the protection of $\mathcal{PT}$ symmetry, each band retains a twofold-degenerate pseudospin subspace, labeled by eigenvalues $\pm 1$, namely, two pseudospin branches. In this respect, the ETM is closer to the charge-nematic metal, whereas this degeneracy is lifted in the polar metal. In the magnetic-quadrupolar metal, the conserved quantity is instead $\hat{s}_z$, so that different FS carry opposite out-of-plane spin polarizations. This can be viewed as the metallic counterpart of altermagnetism\cite{PhysRevB.75.115103}, where spin-split electronic structures arise without a net magnetization. As we shall see below, the ETM is nevertheless distinguished by a tunable pseudospin texture, whose topological structure can be controlled by temperature.

\begin{figure}
\includegraphics[width=0.96\columnwidth]{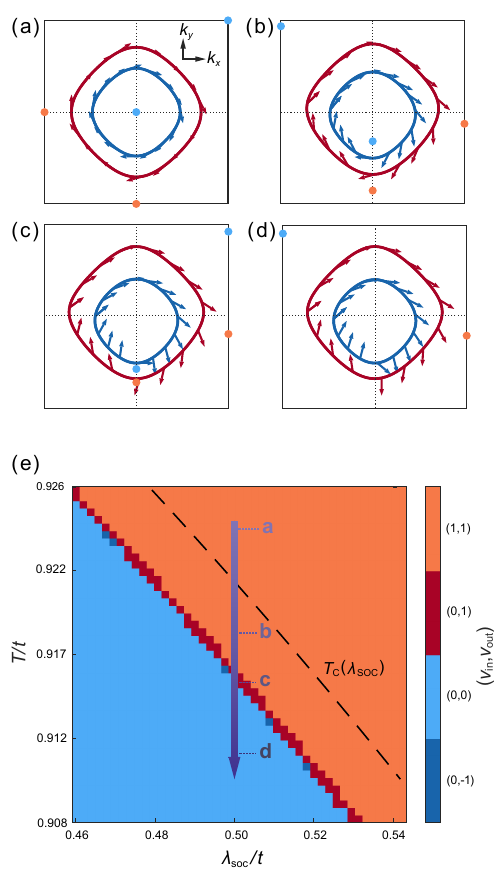}
\caption{The evolution of the topological structure in momentum space. (a)--(d) Pseudospin textures of the ETM for four representative parameter sets, corresponding respectively to the points labeled a--d in panel (e). The textures are shown for the pseudospin branch with eigenvalue $+1$. The filled blue and orange circles mark the centers of pseudospin vortices carrying vorticity $\pm 1$, respectively. (e) Topological phase diagram at fixed $g_{\rm T}=1.5t$, with the phases classified by the pseudospin winding numbers on the FS as temperature and SOC strength are varied.
}
\label{fig3}
\end{figure}

We further examine the intrinsic structure of the ETM in momentum space. Writing the mean-field Hamiltonian as $H_{\rm MF}(\bm{k}) = \xi_{\rm MF}(\bm{k}) \mathbf{1}_4 + \bm{d}_{\rm MF}(\bm{k}) \cdot \bm{\gamma}$, where the three components of $\bm{\gamma}$ are mutually anticommuting matrices (see Supplemental Material for their explicit form), we define the pseudospin orientations of two branches as the directions of $\pm\bm{d}(\bm{k})$. We track the orientation for a fixed branch in momentum space and extract the associated winding numbers on the inner and outer FS, denoted by $\nu_{\rm in}$ and $\nu_{\rm out}$, respectively. This allows us to construct the topological phase diagram shown in Fig.~\ref{fig3}(e). To illustrate the evolution across this phase diagram, we select four representative points along decreasing temperature and plot the corresponding pseudospin textures on the FS in Figs.~\ref{fig3}(a)--(d), where the centers of the pseudospin vortices and their vorticities are explicitly marked.

Starting from point $a$ in Fig.~\ref{fig3}(e), the system is above $T_{\rm c}$, where time-reversal symmetry pins the pseudospin vortices to the four time-reversal-invariant momenta. Upon cooling below $T_{\rm c}$, these vortex centers are released and evolve continuously in momentum space. Although their positions appear to jump between Figs.~\ref{fig3}(a)--(d), the apparent discontinuity merely reflects the choice of Brillouin-zone representation: points differing by reciprocal lattice vectors are physically equivalent, and the vortex motion is continuous on the torus. As the temperature is further reduced, a pair of vortices with opposite vorticity approaches each other and eventually annihilates.

The evolution of the pseudospin vortices determines the winding topology of the FS, since the winding number is given by the net vorticity enclosed by each FS. A distinct regime appears when the vortex initially located at $\bm{k}=0$ crosses the FS and enters the annular region between the inner and outer FS, causing the two FS to enclose different vortex charges and acquire different winding numbers. As shown in Fig.~\ref{fig3}(e), this regime forms a narrow strip between the $(\nu_{\rm in},\nu_{\rm out})=(1,1)$ and $(0,0)$ phases. It lies entirely within the ETM phase and is characterized by a nonzero winding on only one FS, representing an intermediate topological regime between the two limiting cases.

This stripe-like topological crossover regime is not merely a formal classification. As we show below, it leaves a distinct and experimentally observable imprint on the magnetoelectric response of the ETM.

\subsection{Magnetoelectric response}

\subsubsection{Intrinsic interband contribution}

The antisymmetric magnetoelectric response has long been regarded as the hallmark of toroidal order in systems with localized magnetic moments. In ETM, however, this response originates from the condensation of an itinerant electronic instability rather than a static magnetic texture. Beyond distinguishing the microscopic mechanisms, this perspective provides the appropriate framework for defining a toroidal moment in a metal.

In the two-dimensional SOC metal considered here, the intraband contribution to the dc magnetoelectric tensor vanishes due to an exact cancellation between the two bands, leaving only the following interband contribution:
\begin{align}
\alpha^{\rm inter}_{ij}(T,\theta) = -2 \int_{\bm k}  &\sum_{s < s'} \frac{n_F(E_s) - n_F(E_{s'}) }{(E_s - E_{s'})^2} \nonumber\\&\times\operatorname{Im} \operatorname{Tr}\left(P_s M_i P_{s'} J_j\right).\label{ME_tensor}
\end{align}
Here the $\bm{k}$ dependence has been suppressed for compactness. $E_s$ and $P_s$, with $s=\pm$, denote respectively the dispersions and projection operators of the two ETM bands, while $J_j$ is the current operator evaluated in the ETM state. In this work we restrict ourselves to the spin contribution to the magnetic moment, $M_i=-\mu_{\rm B}\tau_0\sigma_i$. The resulting magnetoelectric tensor depends explicitly on both the magnitude $\eta$ and the in-plane orientation $\theta$ of the toroidal vector. Its temperature dependence enters through the Fermi distribution functions and, more importantly, through the temperature evolution of $\eta$.

In an Edelstein-type magnetoelectric response, inversion symmetry breaking alone is sufficient, and the dominant contribution is typically an extrinsic intraband effect\cite{EDELSTEIN1990233,PhysRevB.67.033104,PhysRevLett.113.157201,PhysRevLett.93.176601,PhysRevLett.96.186605}. It originates from the electric-field-driven displacement of the FS and therefore relies on a nonequilibrium relaxation process, resulting in a response proportional to the relaxation time $\tau$. In contrast, the leading contribution in the ETM is an intrinsic interband response that requires the simultaneous breaking of $\mathcal{P}$ and $\mathcal{T}$ symmetries. It does not involve Fermi-surface drift or impurity scattering, but instead arises from quantum-coherent virtual transitions between bands, as reflected by the second-order perturbative structure of Eq.~\ref{ME_tensor}.

A residual intraband background, if present, can be separated from the intrinsic interband response by reversing the orientation of the ETM order parameter. Since time reversal maps $\bm{\eta}$ to $-\bm{\eta}$, the intrinsic interband contribution is odd under this operation, $\alpha_{ij}^{\rm inter}(\theta+\pi)=-\alpha_{ij}^{\rm inter}(\theta)$. By contrast, contributions not associated with the electric-toroidal condensate are even under this reversal. The intrinsic component can therefore be extracted by antisymmetrizing the measured response, $\alpha^{\rm inter}(\theta)=\left[\alpha^{\rm full}(\theta)-\alpha^{\rm full}(\theta+\pi)\right]/2$. The orientation of $\bm{\eta}$ can in principle be controlled by crossed electromagnetic fields through the symmetry-allowed coupling $\Delta F\sim-\bm{\eta}\cdot(\bm{E}\times\bm{B})$\cite{PhysRevB.88.024414}, providing a route to isolating the magnetoelectric signal in the ETM.

For metallic systems, where strong electrostatic screening makes current-driven measurements more natural, the magnetoelectric response is more conveniently expressed as a current-induced magnetization, $M_i=\chi_{ij}J_j$, with $\chi_{ij}=\alpha_{ik}\rho_{kj}$. In the present model, only the out-of-plane magnetization component contributes to this response. Furthermore, the $\mathcal{PT}$-symmetric ETM has vanishing Berry curvature and therefore no Hall conductivity, reducing the relation to $\chi_{zj}(\theta)=\alpha_{zj}(\theta)\rho_{jj}(\theta)$, where $\rho_{jj}$ denotes the longitudinal resistivity. Since the condensate orientation $\theta$ selects a particular direction and breaks the underlying fourfold rotational symmetry, the longitudinal resistivity becomes generally anisotropic, $\rho_{xx}\neq\rho_{yy}$, and acquires an intrinsic dependence on $\theta$.

\begin{figure*}
\centering
\includegraphics[width=0.96\textwidth]{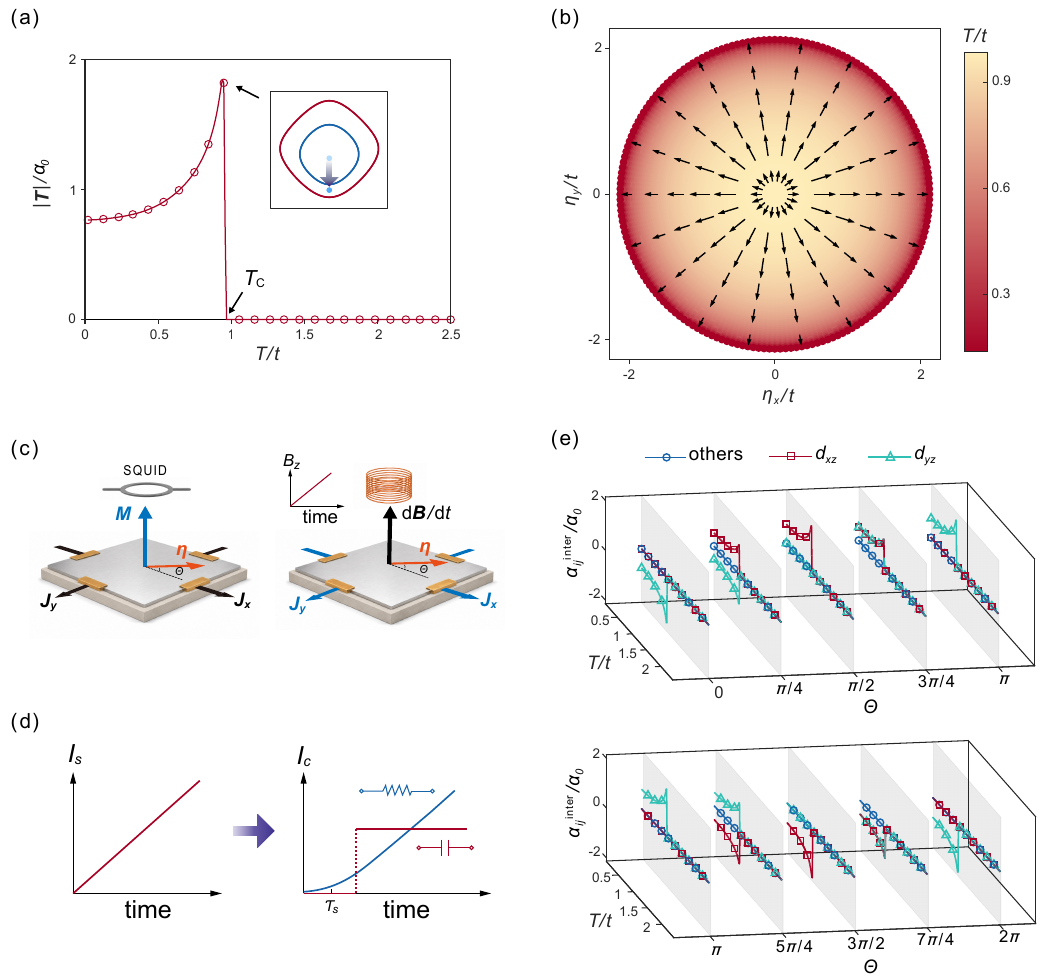}
\caption{Magnetoelectric response of the ETM.
(a) Temperature dependence of the toroidal-moment magnitude. A sharp enhancement below $T_{\rm c}$ signals a strong magnetoelectric response tied to the vortex center crossing the FS. The response is normalized by $\alpha_0=10^{-2}e\mu_{\rm B}/t$ in units with $a=\hbar=1$, where $a$ is the lattice constant.
(b) Evolution of the TM, shown by black arrows in the polar space of the magnitude and angle of the toroidal vector $\bm{\eta}$; color denotes temperature.
(c) Schematic ETM magnetoelectric response: Black arrows denote the perturbations, while blue arrows indicate the induced responses. Red arrow marks the in-plane toroidal vector of the present model. (d) Charge current induced by injecting a $z$-polarized spin current that increases linearly in time. The blue curve denotes the inverse-Edelstein-induced charge current, which grows linearly for times much longer than $\tau_s$ and is therefore conductance-like in the time domain. By contrast, the red curve represents the ETM magnetoelectric response, where the induced charge current remains constant, exhibiting a capacitance-like temporal character. (e) Magnetoelectric quadrupolar components projected onto $d$-wave channels for different in-plane toroidal orientations $\theta$.
}
\label{fig4}
\end{figure*}

Further consider the following set of linear-response relations connected by Onsager reciprocity:
\begin{align}
&M_i(\omega,\theta) = \alpha_{ij}^{\rm inter}(\omega,\theta) E_j(\omega),\\
&J_i(\omega,\theta) = - i \omega \alpha_{ji}^{\rm inter}(\omega,\theta+\pi) B_j(\omega).\label{ME_inverse}
\end{align}
According to Eq.~\ref{ME_inverse}, in the dc limit, the interband magnetoelectric response also gives rise to a charge-pumping effect driven by a linearly varying magnetic field. Specifically, a magnetic field with a constant time derivative induces a dc current described by $J_i(\theta)=\alpha_{ji}^{\rm inter}(\theta+\pi)\dot{B}_j$. This response can be realized, for example, through spintronic approaches, where a spin chemical potential $\mu^s_j$ polarized along the $j$ direction acts as an effective magnetic field $B^{\rm eff}_j$. Neglecting spin diffusion, the spin chemical potential follows the spin continuity equation $\mathrm{d}\mu^s_j/\mathrm{d}t=-\mu^s_j/\tau_s+\kappa I_s$, where $\tau_s$ and $I_s$ denote the spin-relaxation time and injected spin current, respectively. Consequently, a linearly increasing spin current injection generates a constant $\dot{B}_j$ and thereby drives a dc charge current, as illustrated in Fig.~\ref{fig4}(d). This effect should be distinguished from the conventional inverse Edelstein effect\cite{PhysRevLett.112.096601,PhysRevLett.119.256801,PhysRevB.108.144430,Ganichev_2002,Lesne_2016}, where the charge current is proportional to the spin accumulation itself, $\bm{J}\sim\bm{\mu}^s$, rather than to its time derivative. In this sense, the inverse Edelstein effect is a conductive-type response, analogous to a current generated by an applied bias, whereas the interband magnetoelectric response found here is capacitive in character, analogous to a current generated by the time derivative of a bias.

For the present model, we evaluate the temperature evolution of the magnetoelectric response tensor. The scalar component vanishes, $\alpha^{(0)}\sim \operatorname{Tr}\alpha^{\rm inter}=0$, while the antisymmetric component, defined through $\alpha^{(1)}_{ij}=\epsilon_{ijk}T_k$, is characterized by a $\mathcal{P}$-odd and $\mathcal{T}$-odd vector $\bm{T}$. Its magnitude as a function of temperature is shown in Fig.~\ref{fig4}(a), and the full trajectory of $\bm{T}$ is displayed in Fig.~\ref{fig4}(b), revealing its direct correspondence with the condensed toroidal vector $\bm{\eta}$. We further decompose the symmetric traceless part of $\alpha^{\rm inter}$ into the five real $d$-wave components, as shown in Fig.~\ref{fig4}(e). Among them, only the $d_{xz}$ and $d_{yz}$ components are nonzero, exhibiting an alternating angular dependence upon rotating the toroidal vector, namely $\alpha^{{\rm inter},d_{xz}}\sim\sin\theta$ and $\alpha^{{\rm inter},d_{yz}}\sim-\cos\theta$. Combining these pieces, the complete interband magnetoelectric response takes the following form, with the corresponding schematic illustrated in Fig.~\ref{fig4}(c):
\begin{align}
&M_z\sim-\rho_{xx}J_x\sin\theta+\rho_{yy}J_y\cos\theta,\\
&\begin{pmatrix}J_x\\J_y\end{pmatrix}\sim\begin{pmatrix}\dot{B}_z\sin\theta \\-\dot{B}_z\cos\theta\end{pmatrix}.
\end{align}

For insulators with localized moments, or for metals in which itinerant carriers move in an essentially static toroidal background, the toroidal moment can be defined microscopically from the underlying atomic-scale magnetic configuration. Its connection to the antisymmetric component of the magnetoelectric response then follows as a consequence of this microscopic multipolar structure. The situation is different in an ETM, where the ordered state is generated by partially occupied itinerant electronic states themselves and the localized-moment construction no longer provides the appropriate bulk characterization. We therefore reserve $\bm{\eta}$ for the spontaneously condensed toroidal vector and define the metallic toroidal moment $\bm{T}$ through the antisymmetric part of the magnetoelectric tensor, $\alpha^{\rm inter}_{ij}=\epsilon_{ijk}T_k$. As shown in Fig.~\ref{fig4}(b), symmetry locks $\bm{T}$ to the direction of $\bm{\eta}$ in the present model. The two quantities are nevertheless physically distinct: $\bm{\eta}$ characterizes the broken-symmetry state, whereas $\bm{T}$ is a response coefficient determined by the reconstructed itinerant electronic structure. Their magnitudes therefore need not be proportional. This distinction is particularly clear in Fig.~\ref{fig4}(a): although the toroidal vector develops continuously below $T_{\rm c}$, $|\bm{T}|$ exhibits a pronounced nonmonotonic enhancement when a pseudospin vortex traverses the FS, before decreasing again at lower temperatures. The temperature evolution of $\bm{T}$ thus contains information beyond the amplitude of the toroidal condensate and provides a transition-resolved signature of the ETM.

The full interband magnetoelectric tensor also contains a symmetric traceless component in addition to its antisymmetric part. This component defines a magnetoelectric quadrupolar response and should be distinguished from the toroidal moment $\bm{T}$. Such a response is not implied by the presence of a static toroidal background alone. In the present ETM, it arises intrinsically from the itinerant band structure reconstructed by $\bm{\eta}$ and evolves together with the electronic phase transition. The ETM therefore supports intertwined toroidal and quadrupolar magnetoelectric responses, which originate from the same condensate but occupy distinct tensor channels.

\subsubsection{Response enhancement}

Both the antisymmetric and symmetric components of the magnetoelectric response matrix exhibit a pronounced enhancement peak in the ordered phase. On the other hand, Fig.~\ref{fig3}(e) shows that, upon cooling below $T_{\rm c}$, the winding numbers carried by the two FS undergo a sequence of changes from $\nu_{\rm in}=\nu_{\rm out}$ to $\nu_{\rm in}\ne\nu_{\rm out}$, and eventually back to $\nu_{\rm in}=\nu_{\rm out}$. Remarkably, the magnetoelectric peak occurs precisely within the narrow topological regime where $\nu_{\rm in}\ne\nu_{\rm out}$, namely when a vortex center crosses one of the FS. To uncover the connection between these two seemingly unrelated phenomena, we now isolate the contribution of a single vortex; a more complete derivation is provided in the Supplemental Material. The contribution of an individual vortex to the interband magnetoelectric response tensor can be approximated as:
\begin{equation}
\alpha^{\rm inter}_{zj}(T) \approx \alpha_{zj}^{\rm reg}(T) + \mathcal{F}_j(T) \left(\nu_{\rm out}^T - \nu_{\rm in}^T\right),\label{EM_tensor_vortex}
\end{equation}
where $\alpha_{zj}^{\rm reg}$ denotes the contribution from the non-vortex background and $\nu_{\rm out}^T-\nu_{\rm in}^T\equiv \sum_a w_a P_a(T)$ denotes the thermally averaged difference between the winding numbers of the outer and inner FS. Here $w_a$ is the vorticity of the $a$-th vortex, and $P_a(T) = \int_{-\infty}^{+\infty} \mathrm{d}\epsilon \, g_T(\epsilon) \, \bm{1}_{\Omega_\epsilon}(\bm{k}_a)$. We have introduced $g_T(\epsilon)=-\partial n_F(\epsilon)/\partial \epsilon$ and the energy-dependent momentum-space region $\Omega_\epsilon=\{\bm{k}\mid E_-(\bm{k})<\epsilon<E_+(\bm{k})\}$. At $\epsilon=0$, $\Omega_\epsilon$ reduces to the momentum-space annulus enclosed between the two FS.

The meaning of Eq.~\ref{EM_tensor_vortex} is twofold. Once a vortex moves into the momentum-space region $\omega_{\epsilon\sim0}$ between the two FS, it becomes activated and contributes to the magnetoelectric response. At finite temperature, this region is thermally broadened and may be viewed as an annulus with thermally smeared boundaries. The factor $\mathcal{F}_j$ characterizes the contribution of an activated vortex to the magnetoelectric response tensor. It is a topology-independent nonuniversal factor, which depends on the relative position between the vortex center and the FS, as well as on the geometry of the FS. in addition to this, the factor $\nu_{\rm out}^T-\nu_{\rm in}^T$ measures the net vorticity contained in the momentum-space region $\Omega_{\epsilon\approx0}$. It is not quantized and can vary continuously with temperature. In the zero-temperature limit, it reduces to the quantized winding-number difference $\nu_{\rm out}-\nu_{\rm in}=0,\pm 1$. Taken together, these observations show that the vortex contribution to the magnetoelectric response tensor is strongly activated when a net vorticity is present between the two FS. At finite temperature, however, this contribution is not directly proportional to $\nu_{\rm out}-\nu_{\rm in}$, but rather to its thermally broadened, temperature-dependent, noninteger counterpart.

\begin{figure}
\includegraphics[width=0.85\columnwidth]{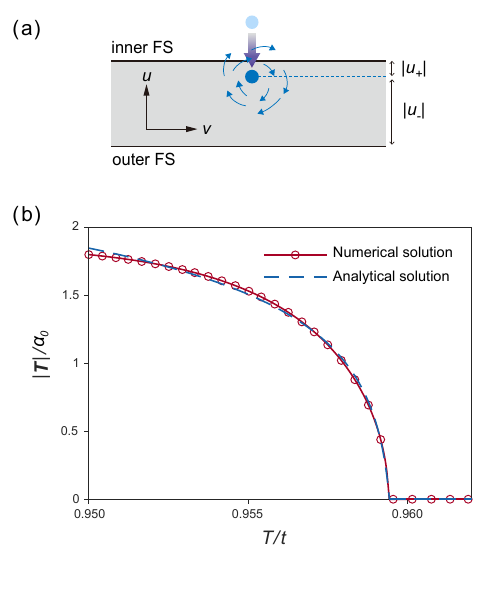}
\caption{(a) Schematic illustration of the local model in momentum space. (b) Comparison between the momentum-space local model and numerical results. In the local model, we consider only the contribution at the FS, neglect the Fermi-surface anisotropy, and adopt the approximation $u_+-u_-\sim\eta\sim\sqrt{T_{\rm c}-T}$}
\label{fig5}
\end{figure}

To provide a more transparent geometric intuition for this result, we now consider a local model in momentum space. We zoom in on the vicinity of a vortex center carrying vorticity $w=\pm 1$. For a given momentum-space region $\Omega_\epsilon$, we ask how the relative position between the vortex center and this momentum region controls the strength of the magnetoelectric response. To this end, we introduce a local Cartesian coordinate system $(u,v)$ centered at the vortex. Since the band dispersion is smooth near the vortex center, the two constant-energy contours that form the boundaries $\partial\Omega_\epsilon$ of the region $\Omega_\epsilon$ can always be approximated locally by two parallel straight lines. We choose the $v$ axis along the tangent direction of these contours, and the $u$ axis perpendicular to them, pointing from the inner boundary to the outer boundary. In this local coordinate system, the region $\Omega_\epsilon$ reduces to an infinitely long strip, $\left\{ u_-(\epsilon)<u<u_+(\epsilon),\, v\in(-\infty,+\infty)\right\}$, as shown in Fig.~\ref{fig5}(a).
If the vortex lies inside this momentum region, then $u_-<0<u_+$; if it lies outside, the origin is on the same side of both boundaries, so that $0$ is not between $u_-$ and $u_+$.

In these local coordinates, the vortex-dominated contribution arises primarily from the tangential current component, and therefore corresponds to the magnetoelectric tensor component $\alpha_{zv}^{\rm vortex}$. Detailed derivation gives:
\begin{align}
&\alpha_{zv}^{\rm vortex}\sim \int_{-\infty}^{+\infty} d\epsilon\, g_T(\epsilon)\mathcal{K}_v(\epsilon),\\
&\mathcal{K}_v(\epsilon)\sim
\frac{w}{(2\pi)^2s^2}
\ln\left[
\frac{m^2+s^2u_+^2(\epsilon)}
{m^2+s^2u_-^2(\epsilon)}
\right].\label{ME_vortex}
\end{align}
Here $g_T(\epsilon)$ restricts the dominant contribution to the vicinity of the Fermi energy, while the geometric information about the relative position between the vortex and the FS is encoded in $\mathcal{K}_v$. The parameters $s$ and $m$ are local Hamiltonian parameters, whose microscopic values are determined by quantities such as $\lambda_{\rm SOC}$ and $m_0$. Thus $\mathcal{K}_v$ represents the local contribution of a single vortex to the magnetoelectric response for a given momentum-space region $\Omega_\epsilon$.

\begin{figure*}
\centering
\includegraphics[width=0.96\textwidth]{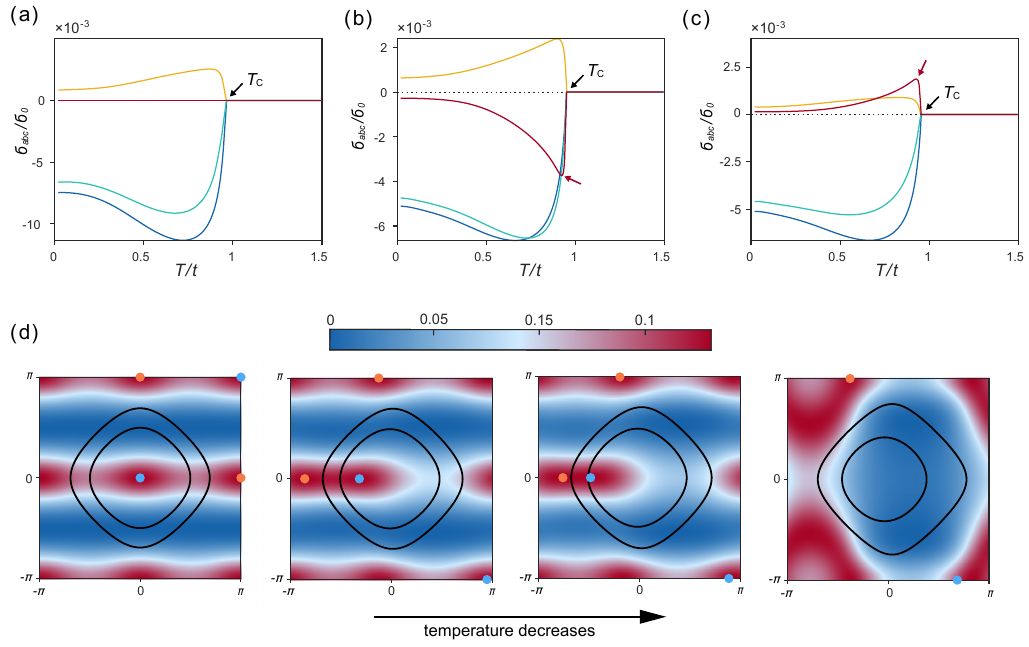}
\caption{Nonlinear conductivity of the ETM with $ \theta=0 $. Panels (a)--(c) show the temperature dependence of $ \sigma_{xxx} $, $ \sigma_{xyy} $, and $ \sigma_{yxy} $, respectively. The blue, red, yellow, and green curves represent the NLD, interQMD, intraQMD, and total contributions, respectively. The conductivity units used for normalization are $ \sigma_0=e^3\tau^2 t $ for the NLD contribution and $ \sigma_0=e^3/t $ for the QMD contributions. In the numerical calculations, we set $ \tau=1 $. (d) The evolution of the interband quantum metric $g_{+-}^{yy}(\bm k)$ in the BZ as the temperature decreases from above $T_{\rm c}$ to below $T_{\rm c}$. Solid circles with different colors indicate the positions of pseudospin vortex centers with different vorticities, while the black curves represent the FS.}
\label{fig6}
\end{figure*}

This expression makes the activation mechanism geometrically transparent. When the vortex is far outside the momentum window, one has $u_+^2\approx u_-^2$, and the logarithm is nearly zero. By contrast, when the vortex crosses the boundary of $\Omega_\epsilon$ and enters the strip, the logarithmic factor becomes sizable and gives the dominant contribution. Moreover, vortices with opposite vorticities contribute with opposite signs. In the distribution of the four vortices in the BZ shown in Figs~\ref{fig3}(a)-(c), two vortices of opposite vorticity remain far from the FS and their contributions nearly cancel. Among the remaining two vortices, one enters the region between the two FS first, thereby producing a pronounced enhancement of the magnetoelectric response.

In the narrow temperature window where a vortex crosses the FS, we compute the contribution of a single vortex to the magnetoelectric response using Eq.~\ref{ME_vortex}, as shown in Fig.~\ref{fig5}(b). Within this parameter regime the momentum-space local model captures well the enhancement of the magnetoelectric response with decreasing temperature, yet it breaks down over a broader temperature range.

\subsection{Nonlinear conductivity}

Having established the magnetoelectric response of the ETM, we now investigate its nonlinear charge transport signatures. The second-order conductivity tensor $ \sigma_{abc} $ can generally be decomposed into four distinct contributions\cite{4z8z-4kch}. The semiclassical contributions consist of the nonlinear Drude (NLD) term and the Berry curvature dipole (BCD) term, which scale as $ \tau^2 $ and $ \tau^1 $, respectively, where $ \tau $ denotes the quasiparticle lifetime. In addition, there exist purely quantum contributions at order $ \tau^0 $, which originate from the quantum geometry of Bloch states and can be further classified into the intraband quantum metric dipole (intraQMD) and interband quantum metric dipole (interQMD) terms.

The nonlinear charge responses of different metallic Landau orders are strongly constrained by their residual spacetime symmetries. In charge-nematic metals and magnetic-quadrupolar metals, the second-order conductivity tensor is completely forbidden, leading to $ \sigma_{abc}=0 $. Both polar metals and ETMs allow a finite NLD contribution; however, their quantum geometric responses are fundamentally different. A polar metal permits a BCD response while forbidding QMD contributions, whereas an ETM, protected by $ \mathcal{PT} $ symmetry, exhibits a vanishing BCD contribution but allows finite QMD responses. Consequently, in the clean flat-band limit, ETMs provide a unique platform among the four classes of metallic Landau orders for realizing intrinsic nonlinear charge transport governed solely by quantum metric and independent of the quasiparticle lifetime.

Since the orientation of the order parameter $ \theta $ determines the direction of the remaining mirror symmetry in the ETM phase, the allowed nonlinear conductivity components are strongly dependent on $ \theta $. In Figs. \ref{fig6}(a)--(c), we present the temperature dependence of the individual contributions to the symmetry-allowed nonlinear conductivity components for $ \theta=0 $. Results for other orientations of the order parameter are provided in the Supplemental Material.

Within the ETM phase, we find that the interQMD contributions to the nonlinear Hall conductivity $ \sigma_{xyy} $ and the mixed component $ \sigma_{yxy} $ exhibit pronounced peaks, as indicated by the red arrows in Figs.~\ref{fig6}(b) and (c). For the present choice of model parameters, the NLD and QMD contributions are comparable in magnitude, with the QMD contribution being slightly smaller, such that the total nonlinear conductivity does not display an apparent peak structure originating from the interQMD contribution. Nevertheless, in systems with weaker band dispersion or nearly flat bands, a pronounced nonlinear conductivity peak is expected to emerge.

As an interband quantum-geometric response, the enhancement of $ \sigma^{\rm interQMD} $ shares the same microscopic origin as that of the magnetoelectric response tensor $ \alpha^{\rm inter} $. Namely, the enhancement occurs when the pseudospin vortex center traverses the FS. This can be understood from the large interband quantum metric weight carried by the vortex. When the vortex approaches the FS and the thermally broadened regions of the two FS overlap in momentum space, the quantum metric weight accumulated between the two FS becomes strongly enhanced, resulting in a significant increase of $ \sigma^{\rm interQMD} $.
For example, for the nonlinear Hall conductivity $ \sigma_{xyy} $, its interQMD part is given by
\begin{equation}
    \sigma^{\rm interQMD}_{xyy}\sim\sum_{m\ne n}\int_{\bm{k}} \frac{v_n^yg^{xy}_{mn}-v^y_ng^{yy}_{mn}}{E_m-E_n}g_T(E_n),\label{sigma_xyy}
\end{equation}
where $v_n^a=\partial_{k_a}E_n$ is the velocity of Bloch electrons, and $g_{mn}^{ab}=\operatorname{Re} \operatorname{Tr}[P_n(\partial_{k_a}P_m)(\partial_{k_b}P_n)]$ denotes the interband quantum metric tensor. In a two-band system, the quantum metric satisfies $g_{+-}^{ab}=g_{-+}^{ba}$. Immediately below the ETM transition, the contribution from $g_{+-}^{xy}$ remains small (see Supplemental Material). Therefore, we focus on the dominant contribution from $g_{+-}^{yy}$ and show its momentum-space distribution in Fig.~\ref{fig6}(d). The quantum metric weight is found to be concentrated around the pseudospin vortices. As the vortex center moves toward the FS, the $g_{+-}^{yy}$ weight accumulated between the two FS is strongly enhanced, providing the dominant contribution to $ \sigma_{xyy}^{\rm interQMD} $. After the annihilation of the two vortices, the quantum metric weight shifts away from the FS, leading to a strong suppression of the interQMD contribution at low temperatures.

\section{Conclusion and discussion}

We have established the electronic toroidal metal as a distinct metallic Landau phase in which the toroidal vector $\bm{\eta}$ emerges spontaneously from a particle-hole instability of an interacting Fermi liquid. Because toroidal electronic bilinears transform identically to a uniform electric current, the physical instability cannot be identified by symmetry alone. The Schur-complement projection removes this noncondensable current component, isolates the genuine toroidal mode, and generates a backflow-corrected electron-boson coupling that is generally nonlocal and retarded.

For itinerant systems, we define the observable toroidal moment $\bm{T}$ through the antisymmetric component of the magnetoelectric tensor, while $\bm{\eta}$ denotes the condensed toroidal vector. In the model studied here, $\bm{T}$ is aligned with $\bm{\eta}$, but its magnitude exhibits a pronounced enhancement governed by the reconstruction of the Fermi-surface pseudospin texture. As pseudospin vortices cross the FS, their winding numbers are redistributed between the inner and outer pockets, strongly enhancing the intrinsic interband magnetoelectric response. By describing the softening, condensation, and response evolution within a single framework, the ETM theory provides a transition-resolved experimental signature beyond the mere observation of a static antisymmetric magnetoelectric coefficient. The ordered phase also supports a symmetric-traceless magnetoelectric response in addition to $\bm{T}$.

The ETM further exhibits a interband quantum-metric contribution to its nonlinear conductivity. The nonlinear Drude term generally remains present; nevertheless, among the four representative spacetime-symmetry classes considered here, the interband quantum-metric contribution distinguishes the ETM and is strongly enhanced by the same pseudospin-vortex reconstruction responsible for the magnetoelectric peak. These results reveal the connection among the toroidal condensate, pseudospin topology, and intrinsic magnetoelectric response.

The present construction can be generalized to other crystalline symmetries, three-dimensional systems, and multiorbital band structures. An immediate extension is to incorporate out-of-plane toroidal channels and determine their competition or coexistence with in-plane order, together with the resulting tensor structure of the magnetoelectric and nonlinear transport responses. 

Beyond mean field, the retarded and nonlocal electron-boson coupling generated by the Schur projection raises the possibility of quantum critical behavior and non-Fermi-liquid dynamics\cite{Lee_2018,Esterlis_2026}. The same low-energy fluctuations may also mediate or reshape superconducting pairing near a toroidal quantum critical point. These questions place the ETM in a broader setting in which toroidal order is not a static background imposed on metallic carriers, but a dynamical Fermi-surface phase whose collective fluctuations can reorganize both the normal state and its neighboring ordered phases.

\section{Acknowledgments}

This work was financially supported by the National Key R\&D Program of China (Grants
No. 2022YFA1403200 and No.2024YFA1613200), National Natural Science Foundation of
China (Grants No. 92265104, No. 12022413), the Basic Research Program of the Chinese
Academy of Sciences Based on Major Scientific Infrastructures (Grant No. JZHKYPT-2021-08), the CASHIPS Directors Fund (Grant No. BJPY2023A09), Anhui Provincial Major S\&T
Project(s202305a12020005), and the High Magnetic Field Laboratory of Anhui Province under
Contract No. AHHM-FX-2020-02

\bibliography{ref.bib}

@article{RevModPhys.81.807,
  title = {Multipolar interactions in $f$-electron systems: The paradigm of actinide dioxides},
  author = {Santini, Paolo and Carretta, Stefano and Amoretti, Giuseppe and Caciuffo, Roberto and Magnani, Nicola and Lander, Gerard H.},
  journal = {Rev. Mod. Phys.},
  volume = {81},
  issue = {2},
  pages = {807--863},
  numpages = {0},
  year = {2009},
  month = {Jun},
  publisher = {American Physical Society},
  doi = {10.1103/RevModPhys.81.807},
  url = {https://link.aps.org/doi/10.1103/RevModPhys.81.807}
}

@article{Pourovskii2025,
    author = {Pourovskii, L. V. and Fiore Mosca, D. and Celiberti, L. and Khmelevskyi, S. and Paramekanti, A. and Franchini, C.},
    title = {Hidden orders in spin–orbit-entangled correlated insulators},
    journal = {Nat. Rev. Mater.},
    volume = {10},
    pages = {674--696},
    year = {2025},
    doi = {10.1038/s41578-025-00824-z}
}

@article{PhysRevB.64.195109,
  title = {{Quantum theory of a nematic Fermi fluid}},
  author = {Oganesyan, Vadim and Kivelson, Steven A. and Fradkin, Eduardo},
  journal = {Phys. Rev. B},
  volume = {64},
  issue = {19},
  pages = {195109},
  numpages = {6},
  year = {2001},
  month = {Oct},
  publisher = {American Physical Society},
  doi = {10.1103/PhysRevB.64.195109},
  url = {https://link.aps.org/doi/10.1103/PhysRevB.64.195109}
}

@article{Chandra_2002, 
 title={Hidden orbital order in the heavy fermion metal $\mathrm{URu_2Si_2}$}, volume={417}, ISSN={1476-4687}, url={http://dx.doi.org/10.1038/nature00795}, DOI={10.1038/nature00795}, number={6891}, journal={Nature}, publisher={Springer Science and Business Media LLC}, author={Chandra, P. and Coleman, P. and Mydosh, J. A. and Tripathi, V.}, year={2002}, pages={831-834} }

@article{PhysRevLett.105.157003,
  title = {Effects of Nematic Fluctuations on the Elastic Properties of Iron Arsenide Superconductors},
  author = {Fernandes, R. M. and VanBebber, L. H. and Bhattacharya, S. and Chandra, P. and Keppens, V. and Mandrus, D. and McGuire, M. A. and Sales, B. C. and Sefat, A. S. and Schmalian, J.},
  journal = {Phys. Rev. Lett.},
  volume = {105},
  issue = {15},
  pages = {157003},
  numpages = {4},
  year = {2010},
  month = {Oct},
  publisher = {American Physical Society},
  doi = {10.1103/PhysRevLett.105.157003},
  url = {https://link.aps.org/doi/10.1103/PhysRevLett.105.157003}
}

@article{PhysRevLett.115.026401,
  title = {Parity-Breaking Phases of Spin-Orbit-Coupled Metals with Gyrotropic, Ferroelectric, and Multipolar Orders},
  author = {Fu, Liang},
  journal = {Phys. Rev. Lett.},
  volume = {115},
  issue = {2},
  pages = {026401},
  numpages = {5},
  year = {2015},
  month = {Jul},
  publisher = {American Physical Society},
  doi = {10.1103/PhysRevLett.115.026401},
  url = {https://link.aps.org/doi/10.1103/PhysRevLett.115.026401}
}

@article{Chu_2010, title={In-Plane Resistivity Anisotropy in an Underdoped Iron Arsenide Superconductor}, volume={329}, ISSN={1095-9203}, url={http://dx.doi.org/10.1126/science.1190482}, DOI={10.1126/science.1190482}, number={5993}, journal={Science}, publisher={American Association for the Advancement of Science (AAAS)}, author={Chu, Jiun-Haw and Analytis, James G. and De Greve, Kristiaan and McMahon, Peter L. and Islam, Zahirul and Yamamoto, Yoshihisa and Fisher, Ian R.}, year={2010}, month=Aug, pages={824-826} }

@article{Kuo_2016, title={{Ubiquitous signatures of nematic quantum criticality in optimally doped Fe-based superconductors}}, volume={352}, ISSN={1095-9203}, url={http://dx.doi.org/10.1126/science.aab0103}, DOI={10.1126/science.aab0103}, number={6288}, journal={Science}, publisher={American Association for the Advancement of Science (AAAS)}, author={Kuo, Hsueh-Hui and Chu, Jiun-Haw and Palmstrom, Johanna C. and Kivelson, Steven A. and Fisher, Ian R.}, year={2016}, month=May, pages={958-962} }

@article{Ronning_2017, title={Electronic in-plane symmetry breaking at field-tuned quantum criticality in $\mathrm{CeRhIn_5}$}, volume={548}, ISSN={1476-4687}, url={http://dx.doi.org/10.1038/nature23315}, DOI={10.1038/nature23315}, number={7667}, journal={Nature}, publisher={Springer Science and Business Media LLC}, author={Ronning, F. and Helm, T. and Shirer, K. R. and Bachmann, M. D. and Balicas, L. and Chan, M. K. and Ramshaw, B. J. and McDonald, R. D. and Balakirev, F. F. and Jaime, M. and Bauer, E. D. and Moll, P. J. W.}, year={2017}, month=Aug, pages={313-317} }

@article{Dubovik_1990, title={Toroid moments in electrodynamics and solid-state physics}, volume={187}, ISSN={0370-1573}, url={http://dx.doi.org/10.1016/0370-1573(90)90042-Z}, DOI={10.1016/0370-1573(90)90042-z}, number={4}, journal={Physics Reports}, publisher={Elsevier BV}, author={Dubovik, V.M. and Tugushev, V.V.}, year={1990}, month=Mar, pages={145-202} }

@book{Landau1980,
  author    = {Landau, L. D. and Lifshitz, E. M.},
  title     = {The Classical Theory of Fields},
  series    = {Course of Theoretical Physics},
  volume    = {2},
  edition   = {4},
  publisher = {Butterworth-Heinemann},
  year      = {1980},
  address   = {Oxford},
  isbn      = {978-0750627689}
}

@article{Spaldin_2008, title={The toroidal moment in condensed-matter physics and its relation to the magnetoelectric effect}, volume={20}, ISSN={1361-648X}, url={http://dx.doi.org/10.1088/0953-8984/20/43/434203}, DOI={10.1088/0953-8984/20/43/434203}, number={43}, journal={Journal of Physics: Condensed Matter}, publisher={IOP Publishing}, author={Spaldin, Nicola A and Fiebig, Manfred and Mostovoy, Maxim}, year={2008}, month=Oct, pages={434203} }

@article{PhysRevLett.102.157203,
  title = {Superexchange-Driven Magnetoelectricity in Magnetic Vortices},
  author = {Delaney, Kris T. and Mostovoy, Maxim and Spaldin, Nicola A.},
  journal = {Phys. Rev. Lett.},
  volume = {102},
  issue = {15},
  pages = {157203},
  numpages = {4},
  year = {2009},
  month = {Apr},
  publisher = {American Physical Society},
  doi = {10.1103/PhysRevLett.102.157203},
  url = {https://link.aps.org/doi/10.1103/PhysRevLett.102.157203}
}

@article{Van_Aken_2007, title={Observation of ferrotoroidic domains}, volume={449}, ISSN={1476-4687}, url={http://dx.doi.org/10.1038/nature06139}, DOI={10.1038/nature06139}, number={7163}, journal={Nature}, publisher={Springer Science and Business Media LLC}, author={Van Aken, Bas B. and Rivera, Jean-Pierre and Schmid, Hans and Fiebig, Manfred}, year={2007}, month=Oct, pages={702-705} }

@article{PhysRevB.82.100408,
  title = {Magnetoelectric $\mathrm{MnPS_3}$ as a candidate for ferrotoroidicity},
  author = {Ressouche, E. and Loire, M. and Simonet, V. and Ballou, R. and Stunault, A. and Wildes, A.},
  journal = {Phys. Rev. B},
  volume = {82},
  issue = {10},
  pages = {100408(R)},
  numpages = {4},
  year = {2010},
  month = {Sep},
  publisher = {American Physical Society},
  doi = {10.1103/PhysRevB.82.100408},
  url = {https://link.aps.org/doi/10.1103/PhysRevB.82.100408}
}

@article{Zimmermann_2014,
  title = {Ferroic nature of magnetic toroidal order},
  author = {Zimmermann, Anne S. and Meier, Dennis and Fiebig, Manfred},
  journal = {Nat. Commun.},
  volume = {5},
  number = {1},
  pages = {4796},
  year = {2014},
  month = {Sep},
  publisher = {Springer Science and Business Media LLC},
  doi = {10.1038/ncomms5796},
  url = {https://doi.org/10.1038/ncomms5796}
}

@article{Ding_2021,
  title = {Field-tunable toroidal moment in a chiral-lattice magnet},
  author = {Ding, Lei and Xu, Xianghan and Jeschke, Harald O. and Bai, Xiaojian and Feng, Erxi and Alemayehu, Admasu Solomon and Kim, Jaewook and Huang, Fei-Ting and Zhang, Qiang and Ding, Xiaxin and Harrison, Neil and Zapf, Vivien and Khomskii, Daniel and Mazin, Igor I. and Cheong, Sang-Wook and Cao, Huibo},
  journal = {Nat. Commun.},
  volume = {12},
  number = {1},
  pages = {1339},
  year = {2021},
  month = {Sep},
  publisher = {Springer Science and Business Media LLC},
  doi = {10.1038/s41467-021-25657-6},
  url = {https://doi.org/10.1038/s41467-021-25657-6}
}

@article{PhysRevB.76.214404,
  title = {Towards a microscopic theory of toroidal moments in bulk periodic crystals},
  author = {Ederer, Claude and Spaldin, Nicola A.},
  journal = {Phys. Rev. B},
  volume = {76},
  issue = {21},
  pages = {214404},
  numpages = {13},
  year = {2007},
  month = {Dec},
  publisher = {American Physical Society},
  doi = {10.1103/PhysRevB.76.214404},
  url = {https://link.aps.org/doi/10.1103/PhysRevB.76.214404}
}

@article{PhysRevB.97.134423,
  title = {Microscopic theory of spin toroidization in periodic crystals},
  author = {Gao, Yang and Vanderbilt, David and Xiao, Di},
  journal = {Phys. Rev. B},
  volume = {97},
  issue = {13},
  pages = {134423},
  numpages = {8},
  year = {2018},
  month = {Apr},
  publisher = {American Physical Society},
  doi = {10.1103/PhysRevB.97.134423},
  url = {https://link.aps.org/doi/10.1103/PhysRevB.97.134423}
}

@article{PhysRevB.98.060402,
  title = {Orbital magnetic quadrupole moment and nonlinear anomalous thermoelectric transport},
  author = {Gao, Yang and Xiao, Di},
  journal = {Phys. Rev. B},
  volume = {98},
  issue = {6},
  pages = {060402(R)},
  numpages = {5},
  year = {2018},
  month = {Aug},
  publisher = {American Physical Society},
  doi = {10.1103/PhysRevB.98.060402},
  url = {https://link.aps.org/doi/10.1103/PhysRevB.98.060402}
}

@article{Bachmann_2020, title={On the absence of stationary currents}, volume={33}, ISSN={1793-6659}, url={http://dx.doi.org/10.1142/S0129055X20600119}, DOI={10.1142/s0129055x20600119}, number={01}, journal={Reviews in Mathematical Physics}, publisher={World Scientific Pub Co Pte Ltd}, author={Bachmann, Sven and Fraas, Martin}, year={2020}, pages={2060011} }

@article{PhysRevLett.123.060601,
  title = {Absence of Energy Currents in an Equilibrium State and Chiral Anomalies},
  author = {Kapustin, Anton and Spodyneiko, Lev},
  journal = {Phys. Rev. Lett.},
  volume = {123},
  issue = {6},
  pages = {060601},
  numpages = {6},
  year = {2019},
  month = {Aug},
  publisher = {American Physical Society},
  doi = {10.1103/PhysRevLett.123.060601},
  url = {https://link.aps.org/doi/10.1103/PhysRevLett.123.060601}
}

@article{PhysRevLett.95.237402,
  title = {{Optical magnetoelectric effect in multiferroic materials: Evidence for a Lorentz force acting on a ray of light}},
  author = {Sawada, Kei and Nagaosa, Naoto},
  journal = {Phys. Rev. Lett.},
  volume = {95},
  issue = {23},
  pages = {237402},
  numpages = {4},
  year = {2005},
  month = {Dec},
  publisher = {American Physical Society},
  doi = {10.1103/PhysRevLett.95.237402},
  url = {https://link.aps.org/doi/10.1103/PhysRevLett.95.237402}
}

@article{PhysRevB.87.014421,
  title = {Symmetry conditions for nonreciprocal light propagation in magnetic crystals},
  author = {Szaller, D\'avid and Bord\'acs, S\'andor and K\'ezsm\'arki, Istv\'an},
  journal = {Phys. Rev. B},
  volume = {87},
  issue = {1},
  pages = {014421},
  numpages = {6},
  year = {2013},
  month = {Jan},
  publisher = {American Physical Society},
  doi = {10.1103/PhysRevB.87.014421},
  url = {https://link.aps.org/doi/10.1103/PhysRevB.87.014421}
}

@article{PhysRevLett.106.057403,
  title = {Enhanced Directional Dichroism of Terahertz Light in Resonance with Magnetic Excitations of the Multiferroic $\mathrm{Ba_2CoGe_2O_7}$ Oxide Compound},
  author = {K\'ezsm\'arki, I. and Kida, N. and Murakawa, H. and Bord\'acs, S. and Onose, Y. and Tokura, Y.},
  journal = {Phys. Rev. Lett.},
  volume = {106},
  issue = {5},
  pages = {057403},
  numpages = {4},
  year = {2011},
  month = {Feb},
  publisher = {American Physical Society},
  doi = {10.1103/PhysRevLett.106.057403},
  url = {https://link.aps.org/doi/10.1103/PhysRevLett.106.057403}
}

@article{Park_2022,
  title = {Nonreciprocal directional dichroism at telecom wavelengths},
  author = {Park, K. and Yokosuk, M. O. and Goryca, M. and Yang, J. J. and Crooker, S. A. and Cheong, S. -W. and Haule, K. and Vanderbilt, D. and Kim, H. -S. and Musfeldt, J. L.},
  journal = {npj Quantum Mater.},
  volume = {7},
  number = {1},
  pages = {43},
  year = {2022},
  month = {Apr},
  publisher = {Springer Science and Business Media LLC},
  doi = {10.1038/s41535-022-00438-6},
  url = {https://doi.org/10.1038/s41535-022-00438-6}
}

@article{PhysRevB.108.L201405,
  title = {Intrinsic nonlinear conductivities induced by the quantum metric},
  author = {Das, Kamal and Lahiri, Shibalik and Atencia, Rhonald Burgos and Culcer, Dimitrie and Agarwal, Amit},
  journal = {Phys. Rev. B},
  volume = {108},
  issue = {20},
  pages = {L201405},
  numpages = {6},
  year = {2023},
  month = {Nov},
  publisher = {American Physical Society},
  doi = {10.1103/PhysRevB.108.L201405},
  url = {https://link.aps.org/doi/10.1103/PhysRevB.108.L201405}
}

@article{4z8z-4kch,
  title = {{Quantum geometric origin of the intrinsic nonlinear Hall effect}},
  author = {Ulrich, Yannis and Mitscherling, Johannes and Classen, Laura and Schnyder, Andreas P.},
  journal = {Phys. Rev. B},
  volume = {113},
  issue = {20},
  pages = {L201107},
  numpages = {9},
  year = {2026},
  month = {May},
  publisher = {American Physical Society},
  doi = {10.1103/4z8z-4kch},
  url = {https://link.aps.org/doi/10.1103/4z8z-4kch}
}

@misc{Shibata2026,
  author = {Shibata, Junya},
  title = {{Second-order dc conductivity in the velocity-gauge Keldysh formalism: gauge-invariant decomposition into nonlinear Drude, Berry-curvature-dipole, and quantum-metric responses}},
  year = {2026},
  eprint = {2606.22359},
  archivePrefix = {arXiv}
}

@article{Samokhin_2022, title={On the effective models of spin-orbit coupling in a two-dimensional electron gas}, volume={437}, ISSN={0003-4916}, url={http://dx.doi.org/10.1016/j.aop.2021.168710}, DOI={10.1016/j.aop.2021.168710}, journal={Annals of Physics}, publisher={Elsevier BV}, author={Samokhin, K.V.}, year={2022}, month=Feb, pages={168710} }

@article{PhysRevLett.108.147003,
  title = {{Topological superconductivity in bilayer Rashba system}},
  author = {Nakosai, Sho and Tanaka, Yukio and Nagaosa, Naoto},
  journal = {Phys. Rev. Lett.},
  volume = {108},
  issue = {14},
  pages = {147003},
  numpages = {5},
  year = {2012},
  month = {Apr},
  publisher = {American Physical Society},
  doi = {10.1103/PhysRevLett.108.147003},
  url = {https://link.aps.org/doi/10.1103/PhysRevLett.108.147003}
}

@article{PhysRevResearch.3.023204,
  title = {Two scenarios for superconductivity in $\mathrm{CeRh_2As_2}$},
  author = {M\"ockli, David and Ramires, Aline},
  journal = {Phys. Rev. Res.},
  volume = {3},
  issue = {2},
  pages = {023204},
  numpages = {6},
  year = {2021},
  month = {Jun},
  publisher = {American Physical Society},
  doi = {10.1103/PhysRevResearch.3.023204},
  url = {https://link.aps.org/doi/10.1103/PhysRevResearch.3.023204}
}

@article{PhysRevB.70.155110,
  title = {Formation of an electronic nematic phase in interacting fermion systems},
  author = {Khavkine, Igor and Chung, Chung-Hou and Oganesyan, Vadim and Kee, Hae-Young},
  journal = {Phys. Rev. B},
  volume = {70},
  issue = {15},
  pages = {155110},
  numpages = {6},
  year = {2004},
  month = {Oct},
  publisher = {American Physical Society},
  doi = {10.1103/PhysRevB.70.155110},
  url = {https://link.aps.org/doi/10.1103/PhysRevB.70.155110}
}

@article{Borzi_2007,
    author = {Borzi, R. A. and Grigera, S. A. and Farrell, J. and Perry, R. S. and Lister, S. J. S. and Lee, S. L. and Tennant, D. A. and Maeno, Y. and Mackenzie, A. P.},
    title = {Formation of a Nematic Fluid at High Fields in $\mathrm{Sr_3Ru_2O_7}$},
    journal = {Science},
    volume = {315},
    number = {5809},
    pages = {214--217},
    year = {2007},
    month = jan,
    issn = {1095-9203},
    doi = {10.1126/science.1134796},
    url = {http://dx.doi.org/10.1126/science.1134796},
    publisher = {American Association for the Advancement of Science (AAAS)}
}

@article{Lawler_2010,
    author = {Lawler, M. J. and Fujita, K. and Lee, Jhinhwan and Schmidt, A. R. and Kohsaka, Y. and Kim, Chung Koo and Eisaki, H. and Uchida, S. and Davis, J. C. and Sethna, J. P. and Kim, Eun-Ah},
    title = {{Intra-unit-cell electronic nematicity of the high-$T_{\rm c}$ copper-oxide pseudogap states}},
    journal = {Nature},
    volume = {466},
    number = {7304},
    pages = {347--351},
    year = {2010},
    month = jul,
    issn = {1476-4687},
    doi = {10.1038/nature09169},
    url = {http://dx.doi.org/10.1038/nature09169},
    publisher = {Springer Science and Business Media LLC}
}

@article{PhysRevLett.14.217,
  title = {Symmetry Considerations on Martensitic Transformations: ``Ferroelectric" Metals?{}},
  author = {Anderson, P. W. and Blount, E. I.},
  journal = {Phys. Rev. Lett.},
  volume = {14},
  issue = {7},
  pages = {217--219},
  numpages = {0},
  year = {1965},
  month = {Feb},
  publisher = {American Physical Society},
  doi = {10.1103/PhysRevLett.14.217},
  url = {https://link.aps.org/doi/10.1103/PhysRevLett.14.217}
}

@article{PhysRevLett.124.237601,
  title = {Multiband Quantum Criticality of Polar Metals},
  author = {Volkov, Pavel A. and Chandra, Premala},
  journal = {Phys. Rev. Lett.},
  volume = {124},
  issue = {23},
  pages = {237601},
  numpages = {7},
  year = {2020},
  month = {Jun},
  publisher = {American Physical Society},
  doi = {10.1103/PhysRevLett.124.237601},
  url = {https://link.aps.org/doi/10.1103/PhysRevLett.124.237601}
}

@article{PhysRevB.107.165110,
  title = {Theory of criticality for quantum ferroelectric metals},
  author = {Klein, Avraham and Kozii, Vladyslav and Ruhman, Jonathan and Fernandes, Rafael M.},
  journal = {Phys. Rev. B},
  volume = {107},
  issue = {16},
  pages = {165110},
  numpages = {33},
  year = {2023},
  month = {Apr},
  publisher = {American Physical Society},
  doi = {10.1103/PhysRevB.107.165110},
  url = {https://link.aps.org/doi/10.1103/PhysRevB.107.165110}
}

@article{Fei_2018,
    author = {Fei, Zaiyao and Zhao, Wenjin and Palomaki, Tauno A. and Sun, Bosong and Miller, Moira K. and Zhao, Zhiying and Yan, Jiaqiang and Xu, Xiaodong and Cobden, David H.},
    title = {{Ferroelectric switching of a two-dimensional metal}},
    journal = {Nature},
    volume = {560},
    number = {7718},
    pages = {336--339},
    year = {2018},
    month = jul,
    issn = {1476-4687},
    doi = {10.1038/s41586-018-0336-3},
    url = {http://dx.doi.org/10.1038/s41586-018-0336-3},
    publisher = {Springer Science and Business Media LLC}
}

@article{PhysRevB.102.014422,
  title = {{Giant momentum-dependent spin splitting in centrosymmetric low-$Z$ antiferromagnets}},
  author = {Yuan, Lin-Ding and Wang, Zhi and Luo, Jun-Wei and Rashba, Emmanuel I. and Zunger, Alex},
  journal = {Phys. Rev. B},
  volume = {102},
  issue = {1},
  pages = {014422},
  numpages = {13},
  year = {2020},
  month = {Jul},
  publisher = {American Physical Society},
  doi = {10.1103/PhysRevB.102.014422},
  url = {https://link.aps.org/doi/10.1103/PhysRevB.102.014422}
}

@article{PhysRevX.12.031042,
  title = {Beyond Conventional Ferromagnetism and Antiferromagnetism: A Phase with Nonrelativistic Spin and Crystal Rotation Symmetry},
  author = {\ifmmode \check{S}\else \v{S}\fi{}mejkal, Libor and Sinova, Jairo and Jungwirth, Tomas},
  journal = {Phys. Rev. X},
  volume = {12},
  issue = {3},
  pages = {031042},
  numpages = {16},
  year = {2022},
  month = {Sep},
  publisher = {American Physical Society},
  doi = {10.1103/PhysRevX.12.031042},
  url = {https://link.aps.org/doi/10.1103/PhysRevX.12.031042}
}

@article{PhysRevLett.132.176702,
  title = {Landau Theory of Altermagnetism},
  author = {McClarty, Paul A. and Rau, Jeffrey G.},
  journal = {Phys. Rev. Lett.},
  volume = {132},
  issue = {17},
  pages = {176702},
  numpages = {8},
  year = {2024},
  month = {Apr},
  publisher = {American Physical Society},
  doi = {10.1103/PhysRevLett.132.176702},
  url = {https://link.aps.org/doi/10.1103/PhysRevLett.132.176702}
}

@article{Reichlova_2024,
    author = {Reichlov{\'a}, Helena and Lopes Seeger, Rafael and Gonz{\'a}lez-Hern{\'a}ndez, Rafael and Kounta, Ismaila and Schlitz, Richard and Kriegner, Dominik and Ritzinger, Philipp and Lammel, Michaela and Leivisk{\"a}, Miina and Hellenes, Anna Birk and Olejn{\'i}k, Kamil and Pet{\v{r}}i{\v{c}}ek, V{\'a}clav and Dole{\v{z}}al, Petr and Horak, Lukas and Schmoranzerova, Eva and Badura, Anton{\'i}n and Bertaina, Sylvain and Thomas, Andy and Baltz, Vincent and Michez, Lisa and Sinova, Jairo and Goennenwein, Sebastian T. B. and {\v{S}}mejkal, Libor and Jungwirth, Tom{\'a}{\v{s}}},
    title = {{Observation of a spontaneous anomalous Hall response in the Mn$_5$Si$_3$ d-wave altermagnet candidate}},
    journal = {Nature Communications},
    volume = {15},
    number = {1},
    pages = {4961},
    year = {2024},
    month = jun,
    doi = {10.1038/s41467-024-48493-w},
    url = {https://doi.org/10.1038/s41467-024-48493-w},
    issn = {2041-1723},
    publisher = {Springer Science and Business Media LLC}
}

@article{Jiang_2025, title={A metallic room-temperature d-wave altermagnet}, volume={21}, ISSN={1745-2481}, url={http://dx.doi.org/10.1038/s41567-025-02822-y}, DOI={10.1038/s41567-025-02822-y}, number={5}, journal={Nature Physics}, publisher={Springer Science and Business Media LLC}, author={Jiang, Bei and Hu, Mingzhe and Bai, Jianli and Song, Ziyin and Mu, Chao and Qu, Gexing and Li, Wan and Zhu, Wenliang and Pi, Hanqi and Wei, Zhongxu and Sun, Yu-Jie and Huang, Yaobo and Zheng, Xiquan and Peng, Yingying and He, Lunhua and Li, Shiliang and Luo, Jianlin and Li, Zheng and Chen, Genfu and Li, Hang and Weng, Hongming and Qian, Tian}, year={2025}, month=Mar, pages={754-759} }

@article{EDELSTEIN1990233,
title = {Spin polarization of conduction electrons induced by electric current in two-dimensional asymmetric electron systems},
journal = {Solid State Communications},
volume = {73},
number = {3},
pages = {233-235},
year = {1990},
issn = {0038-1098},
doi = {https://doi.org/10.1016/0038-1098(90)90963-C},
url = {https://www.sciencedirect.com/science/article/pii/003810989090963C},
author = {V.M. Edelstein},
}

@article{PhysRevB.67.033104,
  title = {{Diffuse transport and spin accumulation in a Rashba two-dimensional electron gas}},
  author = {Inoue, Jun-ichiro and Bauer, Gerrit E. W. and Molenkamp, Laurens W.},
  journal = {Phys. Rev. B},
  volume = {67},
  issue = {3},
  pages = {033104},
  numpages = {4},
  year = {2003},
  month = {Jan},
  publisher = {American Physical Society},
  doi = {10.1103/PhysRevB.67.033104},
  url = {https://link.aps.org/doi/10.1103/PhysRevB.67.033104}
}

@article{PhysRevLett.113.157201,
  title = {{Relativistic N\'eel-order fields induced by electrical current in antiferromagnets}},
  author = {\ifmmode \check{Z}\else \v{Z}\fi{}elezn\'y, J. and Gao, H. and V\'yborn\'y, K. and Zemen, J. and Ma\ifmmode \check{s}\else \v{s}\fi{}ek, J. and Manchon, Aur\'elien and Wunderlich, J. and Sinova, Jairo and Jungwirth, T.},
  journal = {Phys. Rev. Lett.},
  volume = {113},
  issue = {15},
  pages = {157201},
  numpages = {5},
  year = {2014},
  month = {Oct},
  publisher = {American Physical Society},
  doi = {10.1103/PhysRevLett.113.157201},
  url = {https://link.aps.org/doi/10.1103/PhysRevLett.113.157201}
}

@article{PhysRevLett.93.176601,
  title = {Current-Induced Spin Polarization in Strained Semiconductors},
  author = {Kato, Y. K. and Myers, R. C. and Gossard, A. C. and Awschalom, D. D.},
  journal = {Phys. Rev. Lett.},
  volume = {93},
  issue = {17},
  pages = {176601},
  numpages = {4},
  year = {2004},
  month = {Oct},
  publisher = {American Physical Society},
  doi = {10.1103/PhysRevLett.93.176601},
  url = {https://link.aps.org/doi/10.1103/PhysRevLett.93.176601}
}

@article{PhysRevLett.96.186605,
  title = {Spectral Dependence of Spin Photocurrent and Current-Induced Spin Polarization in an $\mathrm{InGaAs}/\mathrm{InAlAs}$ Two-Dimensional Electron Gas},
  author = {Yang, C. L. and He, H. T. and Ding, Lu and Cui, L. J. and Zeng, Y. P. and Wang, J. N. and Ge, W. K.},
  journal = {Phys. Rev. Lett.},
  volume = {96},
  issue = {18},
  pages = {186605},
  numpages = {4},
  year = {2006},
  month = {May},
  publisher = {American Physical Society},
  doi = {10.1103/PhysRevLett.96.186605},
  url = {https://link.aps.org/doi/10.1103/PhysRevLett.96.186605}
}

@article{PhysRevB.88.024414,
  title = {Controlling toroidal moments by crossed electric and magnetic fields},
  author = {Baum, M. and Schmalzl, K. and Steffens, P. and Hiess, A. and Regnault, L. P. and Meven, M. and Becker, P. and Bohat\'y, L. and Braden, M.},
  journal = {Phys. Rev. B},
  volume = {88},
  issue = {2},
  pages = {024414},
  numpages = {5},
  year = {2013},
  month = {Jul},
  publisher = {American Physical Society},
  doi = {10.1103/PhysRevB.88.024414},
  url = {https://link.aps.org/doi/10.1103/PhysRevB.88.024414}
}

@article{PhysRevLett.112.096601,
  title = {{Microscopic theory of the inverse Edelstein effect}},
  author = {Shen, Ka and Vignale, G. and Raimondi, R.},
  journal = {Phys. Rev. Lett.},
  volume = {112},
  issue = {9},
  pages = {096601},
  numpages = {5},
  year = {2014},
  month = {Mar},
  publisher = {American Physical Society},
  doi = {10.1103/PhysRevLett.112.096601},
  url = {https://link.aps.org/doi/10.1103/PhysRevLett.112.096601}
}

@article{PhysRevLett.119.256801,
  title = {Theory of the Spin Galvanic Effect at Oxide Interfaces},
  author = {Seibold, G\"otz and Caprara, Sergio and Grilli, Marco and Raimondi, Roberto},
  journal = {Phys. Rev. Lett.},
  volume = {119},
  issue = {25},
  pages = {256801},
  numpages = {5},
  year = {2017},
  month = {Dec},
  publisher = {American Physical Society},
  doi = {10.1103/PhysRevLett.119.256801},
  url = {https://link.aps.org/doi/10.1103/PhysRevLett.119.256801}
}

@article{PhysRevB.108.144430,
  title = {{Theory of inverse Rashba-Edelstein effect induced by spin pumping into a two-dimensional electron gas}},
  author = {Yama, M. and Matsuo, M. and Kato, T.},
  journal = {Phys. Rev. B},
  volume = {108},
  issue = {14},
  pages = {144430},
  numpages = {15},
  year = {2023},
  month = {Oct},
  publisher = {American Physical Society},
  doi = {10.1103/PhysRevB.108.144430},
  url = {https://link.aps.org/doi/10.1103/PhysRevB.108.144430}
}

@article{Ganichev_2002,
    author = {Ganichev, S. D. and Ivchenko, E. L. and Bel'kov, V. V. and Tarasenko, S. A. and Sollinger, M. and Weiss, D. and Wegscheider, W. and Prettl, W.},
    title = {{Spin-galvanic effect}},
    journal = {Nature},
    volume = {417},
    number = {6885},
    pages = {153--156},
    year = {2002},
    month = may,
    issn = {1476-4687},
    doi = {10.1038/417153a},
    url = {http://dx.doi.org/10.1038/417153a},
    publisher = {Springer Science and Business Media LLC}
}

@article{Lesne_2016,
    author = {Lesne, E. and Fu, Yu and Oyarzun, S. and Rojas-Sánchez, J. C. and Vaz, D. C. and Naganuma, H. and Sicoli, G. and Attané, J.-P. and Jamet, M. and Jacquet, E. and George, J.-M. and Barthélémy, A. and Jaffrès, H. and Fert, A. and Bibes, M. and Vila, L.},
    title = {{Highly efficient and tunable spin-to-charge conversion through Rashba coupling at oxide interfaces}},
    journal = {Nature Materials},
    volume = {15},
    number = {12},
    pages = {1261--1266},
    year = {2016},
    month = aug,
    doi = {10.1038/nmat4726},
    url = {http://dx.doi.org/10.1038/nmat4726},
    publisher = {Springer Science and Business Media LLC}
}

@article{Lee_2018, title={{Recent developments in non-Fermi liquid theory}}, volume={9}, ISSN={1947-5462}, url={http://dx.doi.org/10.1146/annurev-conmatphys-031016-025531}, DOI={10.1146/annurev-conmatphys-031016-025531}, number={1}, journal={Annual Review of Condensed Matter Physics}, publisher={Annual Reviews}, author={Lee, Sung-Sik}, year={2018}, month=Mar, pages={227-244} }

@article{Esterlis_2026,
    author = {Esterlis, Ilya and Schmalian, Jörg},
    title = {{Quantum critical Eliashberg theory}},
    journal = {Annual Review of Condensed Matter Physics},
    volume = {17},
    number = {1},
    pages = {419--448},
    year = {2026},
    month = mar,
    doi = {10.1146/annurev-conmatphys-032822-042856},
    url = {https://doi.org/10.1146/annurev-conmatphys-032822-042856},
    publisher = {Annual Reviews},
    issn = {1947-5454}
}

@article{PhysRevB.75.115103,
  title = {{Fermi liquid instabilities in the spin channel}},
  author = {Wu, Congjun and Sun, Kai and Fradkin, Eduardo and Zhang, Shou-Cheng},
  journal = {Phys. Rev. B},
  volume = {75},
  issue = {11},
  pages = {115103},
  numpages = {25},
  year = {2007},
  month = {Mar},
  publisher = {American Physical Society},
  doi = {10.1103/PhysRevB.75.115103},
  url = {https://link.aps.org/doi/10.1103/PhysRevB.75.115103}
}

\end{document}